\documentclass[twocol]{ametsocV6}
\usepackage[T1]{fontenc}
\usepackage{float}
\usepackage{overpic}
\usepackage{amsmath,amssymb,amsfonts,bm}
\usepackage{natbib}
\usepackage{url}
\makeatletter
\def\amsbb{\use@mathgroup \M@U \symAMSb}
\makeatother
\usepackage[bbgreekl]{mathbbol}
\newcommand{\xdownarrow}[1]{{\left\downarrow\vbox to #1{}\right.\kern-\nulldelimiterspace}}
\newcommand{\bra}[1]{\langle #1\rangle}

\graphicspath{{./Figs/}{./matlab/}}

\newcommand\BibTeX{{\rmfamily B\kern-.05em \textsc{i\kern-.025em b}\kern-.08em
T\kern-.1667em\lower.7ex\hbox{E}\kern-.125emX}}

\renewcommand{\vec}[1]{\ensuremath{\mbox{\boldmath$#1$}}}

\newcommand{\Tab}[1]{Table~\ref{#1}}
\newcommand{\Fig}[1]{Fig.~\ref{#1}}
\newcommand{\Figs}[2]{Figs.~\ref{#1} and \ref{#2}}

\newcommand{\Eq}[1]{Eq.~(\ref{#1})}
\newcommand{\Eqs}[2]{Eqs.~(\ref{#1}) and~(\ref{#2})}
\newcommand{\Eqss}[2]{equations~(\ref{#1})--(\ref{#2})}
\newcommand{\um}{\,\mu{\rm m}}
\newcommand{\cm}{\,\rm cm}
\newcommand{\m}{\,\rm m}
\newcommand{\s}{\,\rm s}
\newcommand{\SKEW}{{\rm skew \ }}
\newcommand{\KURT}{{\rm kurt \ }}

\def\blue{\textcolor{black}}

\newcommand{\nps}{\xi}

\title{Collision fluctuations of lucky droplets with superdroplets}
\date{\today,~ $ $Revision: 1.801 $ $}

\authors{
Xiang-Yu Li\aff{a, b, c, d}\correspondingauthor{Xiang-Yu Li, xiang.yu.li@su.se,
\today,~ $ $Revision: 1.801 $ $},
Bernhard Mehlig\aff{e},
Gunilla Svensson\aff{a, c},
Axel Brandenburg\aff{b, d, f},
Nils E.\ L.\ Haugen\aff{g, h}
}
\affiliation{
\aff{a}Department of Meteorology and Bolin Centre for Climate Research, Stockholm University, Stockholm, Sweden \\
\aff{b}Nordita, KTH Royal Institute of Technology and Stockholm University, 10691 Stockholm, Sweden \\
\aff{c}Swedish e-Science Research Centre, www.e-science.se, Stockholm, Sweden \\
\aff{d}JILA and Laboratory for Atmospheric and Space Physics, University of Colorado, Boulder, CO 80303, USA \\
\aff{e}Department of Physics, Gothenburg University, 41296 Gothenburg, Sweden \\
\aff{f}The Oskar Klein Centre, Department of Astronomy, Stockholm University, AlbaNova, SE-10691 Stockholm, Sweden\\
\aff{g}SINTEF Energy Research, 7465 Trondheim, Norway \\
\aff{h}Division of Energy Science, Lule\aa \, University of Technology, Lule\aa \,  971 87, Sweden \\
}

\abstract{
It was previously shown that the superdroplet algorithm for modeling the
collision-coalescence process can faithfully represent
mean droplet growth in turbulent clouds.
But an open question is how accurately the superdroplet algorithm
accounts for fluctuations in the collisional aggregation process.
Such fluctuations are particularly important in dilute suspensions.
Even in the absence of turbulence, Poisson fluctuations of collision
times in dilute suspensions may result in substantial variations in the growth process,
resulting in a broad distribution of growth times to reach a
certain droplet size.
We quantify the accuracy of the superdroplet algorithm in describing
the fluctuating growth history of a larger droplet that
settles under the effect of gravity in a quiescent fluid and collides with
a dilute suspension of smaller droplets that were initially randomly
distributed in space (`lucky droplet model').
We assess the effect of fluctuations upon the growth history
of the lucky droplet and compute the distribution of cumulative collision times.
The latter is shown to be sensitive enough to detect the subtle increase
of fluctuations associated with collisions between multiple lucky droplets.
The superdroplet algorithm incorporates fluctuations in two distinct ways:
through the random spatial distribution of superdroplets and
through the Monte Carlo collision algorithm involved.
Using specifically designed numerical experiments, we show that \blue{both}
on their own give an accurate representation of fluctuations.
We conclude that the superdroplet algorithm can faithfully
represent fluctuations in the coagulation of droplets
driven by gravity.
}

\begin{document}
\maketitle

\section{Introduction}
\label{Introduction}

Direct numerical simulations (DNS) have become an essential tool
to investigate collisional growth of droplets in turbulence
\citep{onishi2015, saito2018turbulence}.
Here, DNS refers to the realistic modeling of all relevant
processes, which involves not only the use of a realistic viscosity, but also
a realistic modeling of collisions of droplet pairs in phase space.
The most natural and physical way to analyze collisional growth
is to track individual droplets and to record their collisions, one by one.
However, DNS of the collision-coalescence process are very challenging,
not only when a large number of droplets must be tracked,
but also because the flow must be resolved over a large range
of time and length scales.

Over the past few decades, an alternative way of modeling aerosols has
gained popularity.
\cite{zannetti1984new} introduced the concept of ``superparticles, i.e.,
simulation particles representing a cloud of physical particles having
similar characteristics.''
This concept was also used by \cite{paoli2004contrail} in the context
of condensation problems.
The application to coagulation problems was pioneered by
\cite{Dullemond_2008} and \cite{Shima09}, who also
developed a computationally efficient algorithm.
The idea is to combine physical cloud droplets into `superdroplets'.
To gain efficiency, one tracks only superdroplet collisions and
uses a Monte Carlo algorithm \citep{Sok97}
to account for collisions between physical droplets.
This is referred to as ``superdroplet algorithm.''
It is used in both
the meteorological literature
\citep{Shima09, solch2010large, Riechelmann12,Arabas13,
Naumann15,Naumann16,Unterstrasser17, Dziekan17,
li17, li2017effect, li2018cloud, li2018condensational, 
sato2017grid, jaruga2018libcloudph, brdar2018mcsnow,sato2018numerical,seifert2019geometry,
hoffmann2019inhomogeneous, dziekan2019university,
grabowski2019modeling,shima2020predicting, grabowski2020comparison, unterstrasser2020collisional}, 
as well as in the astrophysical literature
\citep{Dullemond_2008, ormel2009dust, zsom2010outcome, Johansen_2012, johansen2015growth, ros2013ice, Dullemond_2014,
kobayashi2019importance,baehr2019concentration, ros2019effect, nesvorny2019trans, yang2020morphological, poon2020formation, li2020dust, li2020coagulation}.
Compared with DNS, the superdroplet algorithm is distinctly more efficient.
It has been shown
to accurately model average properties of droplet growth in turbulent clouds.
\citet{li2017effect} demonstrated, for example, that the mean collision rate obtained using
the superdroplet algorithm agrees with the mean turbulent
collision rate \citep{1955_Saffman} when the droplets are small.

Less is known about how the superdroplet algorithm represents fluctuations
in the collisional aggregation process.
\citet{Dziekan17} compared the results of the
superdroplet algorithm with the predictions of the stochastic coagulation
equation of \cite{gillespie1972stochastic} in the context of
coalescence of droplets settling in a quiescent fluid.
\citet{Dziekan17} concluded that the results of the superdroplet algorithm qualitatively
agree with what \citet{Kos05} called the lucky droplet model (LDM).
To assess the importance of fluctuations,
\citet{Dziekan17} computed the time $t_{10\%}$,
after which 10\% of the droplets have reached a radius of $40\um$.
In agreement with earlier Lagrangian simulations of \cite{onishi2015},
which did not employ the superdroplet algorithm, they found that
the difference in $t_{10\%}$ between their superdroplet simulations
and the stochastic model of \citep{gillespie1972stochastic}
decreases with the square root of the number of
droplets, provided that there are no more than about nine droplets per superdroplet.
The number of droplets in each superdroplet \blue{$i$} is called the multiplicity
\blue{
$\nps_i(t)$.
}
When this number is larger than $9$, they found that
a residual error remains.
We return to this question in the discussion of the present study, where we
tentatively associate their findings with the occurrence of several
large (lucky) droplets that grew from the finite tail of their
initial droplet distribution.
 
The role of fluctuations is particularly important in dilute
systems, where rare extreme events may substantially broaden
the droplet-size distribution.
This is well captured by the LDM, which was first
proposed by \cite{Telford1955} and later numerically addressed by
\cite{twomey1964statistical}, and more recently quantitatively analyzed
by \cite{Kos05}.
The model describes one droplet of $12.6\um$ radius
settling through a dilute suspension of background droplets
with $10\um$ radius.
The collision times between the larger (``lucky'') droplet and
the smaller ones are exponentially distributed, leading to substantial
fluctuations in the growth history of the lucky droplet.
\citet{Wilkinson16} derived analytic expressions for the
cumulative distribution times using large-deviation theory.
\cite{madival2018stochastic} extended the theory of \cite{Kos05}
by considering a more general form of the droplet-size distribution than
just the Poisson distribution. 

The goal of the present study is to investigate how accurately the
superdroplet algorithm represents fluctuations in the collisional growth
history of settling droplets in a quiescent fluid.
Unlike the work of \citet{Dziekan17}, who focused on the
calculation of $t_{10\%}$, we compare here with the distribution of cumulative
collision times, which is the key diagnostics of the LDM.
We record growth histories of the larger droplet in an
ensemble of different realizations of identical smaller droplets that
were initially randomly distributed in a quiescent fluid.
We show that the superdroplet algorithm accurately describes the
fluctuations of growth histories of the lucky droplet in an ensemble of
simulations.
\blue{In its simplest form, the} LDM assumes that the
lucky droplet is large compared to the background droplets, so that the
radius of those smaller droplets can be neglected in the geometrical
collision cross section and velocities of colliding droplets;
\blue{
see Eqs.~(3) and (4) of \cite{Kos05}, for example.
}
Since fluctuations early on in the growth history are most important
\citep{Kos05,Wilkinson16}, this can make a certain difference in
the distribution of the time $T$ it takes for the lucky droplet to
grow to a certain size.
As the small droplets are initially randomly distributed,
their local number density fluctuates.
Consequently, lucky droplets can grow most quickly where the local
number density of small droplets happens to be large.

The remainder of this study is organized as follows.
In section~\ref{sec:method} we describe the superdroplet algorithm and highlight
differences between different implementations used in the literature
\citep{Shima09,Johansen_2012, li17}.
Section~\ref{sec:ldm} summarizes the LDM,
the setup of our superdroplet simulations, and how we
measure fluctuations of growth histories.
Section~\ref{sec:gh} summarizes the results of our superdroplet simulations.
We conclude in section~\ref{sec:conclusion}.

\section{Method}
\label{sec:method}

\subsection{Superdroplet algorithm}
\label{sec:alg}

\begin{table}[b]
\begin{center}  
\caption{\label{tab:1} Definition of variables in superdroplet algorithm.}
\begin{tabular}{ll}
\hline\hline
$n\!\!$ & number density of droplets in the domain \\
$n_{\rm luck}\!\!$ & number density of lucky droplets \\
$N_{\rm s}(t)\!\!$ & number of ``superdroplets'' in the domain \\
$\nps_i(t)\!\!$ & number of droplets in superdroplet $i$ (multiplicity) \!\!\!\! \\
$N_{\rm d}(t)\!\!$ & total number of physical droplets in the domain \\
$N_{\rm real}\!\!$ & number of independent simulations (realizations)\\
\hline\hline
\end{tabular}
\end{center}
\end{table}

Superdroplet algorithms represent several physical droplets by one superdroplet.
All droplets in superdroplet $i$ are assumed to have the same
material density $\rho_{\rm d}$, the same radius $r_i$, the same
velocity $\bm v_i$, and reside in a volume around the
same position $\bm x_i$.
The index $i$ labeling the superdroplets ranges from $1$ to $N_{\rm s}(t_0)$
(Table~\ref{tab:1}), where $t_0$ denotes the initial time.

\begin{figure*}
\begin{center}
\begin{overpic}[width=\textwidth]{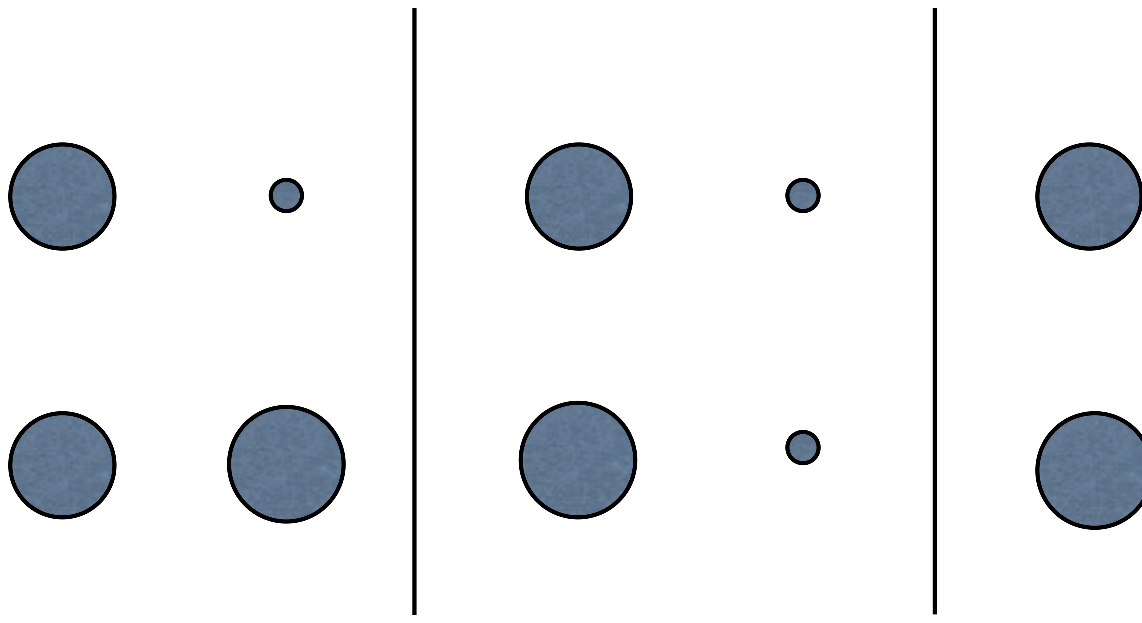}
\put(-5,18){$\xdownarrow{1.3cm}$}
\put(-8,18){$t$}
\put(17,38){(a)}
\put(49,38){(b)}
\put(82,38){(c)}
\put(5,32){$\scriptstyle  \nps_i = 10$}
\put(20,32){$\scriptstyle  \nps_j = 6$}
\put(5,21){$\scriptstyle M_i = 10$}
\put(20,21){$\scriptstyle M_j = 2$}
\put(5,15){$\scriptstyle  \nps_i = 4$}
\put(20,15){$\scriptstyle  \nps_j = 6$}
\put(5,3){$\scriptstyle M_i = 10$}
\put(20,3){$\scriptstyle M_j = 12$}
\put(38,32){$\scriptstyle \nps_i = 6$}
\put(52,32){$\scriptstyle \nps_j = 10$}
\put(38,21){$\scriptstyle M_i = 10$}
\put(52,21){$\scriptstyle M_j = 2$}
\put(38,15){$\scriptstyle  \nps_i = 6$}
\put(52,15){$\scriptstyle  \nps_j = 4$}
\put(38,3){$\scriptstyle M_i = 12$}
\put(52,3){$\scriptstyle M_j = 2$}
\put(71,32){$\scriptstyle  \nps_i = 8$}
\put(86,32){$\scriptstyle  \nps_j = 8$}
\put(71,21){$\scriptstyle M_i = 10$}
\put(86,21){$\scriptstyle M_j = 2$}
\put(71,15){$\scriptstyle  \nps_i = 4$}
\put(86,15){$\scriptstyle  \nps_j = 4$}
\put(71,3){$\scriptstyle M_i = 12$}
\put(86,3){$\scriptstyle M_j = 12$}
\end{overpic}
\end{center}
\caption{\label{fig:2}
Collision outcomes with (a): $\xi_i>\xi_j$,
(b): $\xi_i<\xi_j$, and (c): $\xi_i=\xi_j$ when two superdroplets collide
and droplet collisions occur.
Time increases downward, as indicated by the arrow.
Superdroplet $i$ contains $\nps_i$ large droplets of mass $M_i$,
superdroplet $j$ contains $\nps_j$ small droplets of mass $M_j<M_i$.
}
\end{figure*}

The equation of motion for the position $\bm{x}_i$ and
velocity $\bm{v}_i$ of superdroplet $i$ reads:
\begin{equation}
\label{dxidt}
\frac{{\rm d}\bm{x}_i}{{\rm d}t}=\bm{v}_i
\,,\quad
\frac{{\rm d}\bm{v}_i}{{\rm d}t}=-\frac{\bm{v}_i}{\tau_i}
  + \bm{g}\,.
\end{equation}
Here $\bm{g}$ is the gravitational acceleration,
and the hydrodynamic force is modeled using Stokes law, so that
\begin{equation}
\label{response_time}
  \tau_i=\frac{2}{9}\frac{\rho_{\rm d}}{\rho} \frac{r_i^2}{\nu}
\end{equation}
is the droplet response (or Stokes) time attributed to the superdroplet,
$\nu=10^{-5}\m^2\s^{-1}$ is the viscosity of air,
and $\rho$ is the mass density of the airflow.
Droplets are only subject to gravity and no turbulent airflow is
simulated.

Droplet collisions are represented by collisions of superdroplets
\citep{Shima09,Johansen_2012, li17}, as mentioned above. 
Superdroplets $i$ and $j$ (collision partners) residing inside a grid cell collide
with probability
\begin{equation}
\label{eq:pij}
p_{ij}=\lambda_{ij}\delta t\,,
\end{equation}
where $\delta t$ is the integration time step
\blue{
and $\lambda_{ij}$ is their collision rate.
}
A collision happens when $\eta<p_{ij}$, where $0\leq\eta\leq1$ is a
uniformly distributed random number.
To avoid a probability larger than unity, we limit the integration step
through the condition $\delta t \ll 1/\lambda_{ij}$ \blue{\citep{Johansen_2012, li17}}.
The collision rate is \blue{given by}
\begin{equation}
\lambda_{ij}=\pi\left(r_i+r_j\right)^2|\bm{v}_i-\bm{v}_j|
   \,E_{ij}\,\frac{\nps_{\max}}{\delta x^3}\,,
\label{eq:collision_rate}
\end{equation}
where $E_{ij}$ is the collision efficiency,
$\nps_{\max}=\max(\nps_i,\nps_j)$ is the larger one of the
\blue{multiplicities $\nps_i$ and $\nps_j$
of superdroplets $i$ or $j$ (Table~\ref{tab:1}), and
}
$\delta x^3$ is the volume of the grid cell
closest to the superdroplet.
\blue{
The number density of physical droplets in superdroplet $i$ is then
$n_i=\nps_i/\delta x^3$.
}
Note that \Eq{eq:collision_rate} implies that droplets
having the same velocity ($\bm{v}_i=\bm{v}_j$)
never collide.
This also implies that no collisions are possible between physical
particles within a single superdroplet.
For the purpose of the present study, it suffices to limit ourselves to
the simplest, albeit unrealistic assumption of $E_{ij}=1$, but we also consider
in one case a slightly more realistic quadratic dependence on the radius
of the larger droplet.
To assess the effects of this assumption, we compare
with results where the efficiency increases with droplet radius
\citep{lamb_verlinde_2011}.
Following \cite{Kos05} and \cite{Wilkinson16}, we adopt a simple
power law prescription for the dependence of the efficiency on
the droplet radius.

What happens when two superdroplets collide? 
The collision scheme suggested by \citet{Shima09} amounts to the following rules;
see also \Fig{fig:2} for an illustration.
To ensure mass conservation between superdroplets $i$ and $j$,
when $\nps_j > \nps_i$, which is the case illustrated in
\Fig{fig:2}(b), droplet numbers and masses are updated such that
\begin{align}
\label{eq:updateN}
\nps_i&\to \nps_i\,, \quad \nps_j \to \nps_j-\nps_i\,,\\
M_i&\to M_i + M_j\,,\quad M_j \to M_j\, ,\nonumber
\end{align}
\blue{where $M_i$ and $M_j$ are the droplet masses.}
When $\nps_j < \nps_i$, which is the case shown in \Fig{fig:2}(a),
the update rule is also given by \Eq{eq:updateN}, but with indices
$i$ and $j$ exchanged.
In other words, the number of droplets in the smaller superdroplet
remains unchanged (and their masses are increased), while that in
the larger one is reduced by the amount of droplets that
have collided with all the droplets of the smaller superdroplet
(and their masses remain unchanged).

To ensure momentum conservation during the collision,
the momenta of droplets in the two superdroplets are updated as
\begin{align}
\bm{v}_i   M_i    &\to \bm{v}_i M_i+\bm{v}_j M_j\,, \nonumber \\
\bm{v}_j   M_j    &\to \bm{v}_j M_j\,,
\end{align}
after a collision of superdroplets.

Finally, when $\nps_i = \nps_j$, which is the case described
in \Fig{fig:2}(c), droplet numbers and masses are updated as
\begin{align}
\label{eq:updateN2}
\nps_i&\to \nps_i/2\,, \quad \nps_j \to \nps_j/2\,,\\
M_i&\to M_i + M_j\,,\quad M_j \to M_i + M_j\,,\nonumber
\end{align}
\blue{and} it is then assumed that, when two superdroplets,
each with one or less than one physical droplet, collide,
the superdroplet containing the smaller physical droplet is
collected by the more massive one; it is thus removed from the
computational domain after the collision\blue{, still conserving mass and momentum.} 
We emphasize that \Eq{eq:updateN} does not require $\xi$ to be an integer.
Since we usually specify the initial number density of physical particles,
$\xi$ can be fractional from the beginning.
This is different from the integer treatment of $\xi$
in \cite{Shima09}.

The superdroplet simulations are performed by using the particle modules
of the {\sc Pencil Code} \citep{Collaboration2021}.
The fluid dynamics modules of the code are not utilized here.
To reduce the computational cost and make it linear in the number of superdroplets
per mesh point, $n_{\rm s}(t)$, \citet{Shima09} supposed that each
superdroplet interacts with only one randomly selected superdroplet per time step rather than
allowing collisions with all the other superdroplets in a grid cell
(they still allow multiple coalescence for randomly generated, non-overlapping candidate pairs
in one time step, which is what they referred to as random
permutation technique. This technique was
also adopted by \citet{Dziekan17} and \citet{unterstrasser2020collisional}.
However, this is not used in the {\sc Pencil Code}.
Instead, we allow each superdroplet to collide with all other superdroplets
within one grid cell to maximize the statistical accuracy of the results.
This leads to a computational cost of $\mathcal{O}(n^2_{\rm s}(t))$, which
does not significantly increase the computational cost
because $n_{\rm s}(t)$ is relatively small for cloud-droplet collision
simulations. 
In the {\sc Pencil Code}, collisions
between particles residing within a given grid cell are evaluated by the same
processor which is also evaluating the equations of that grid cell. Due to
this, together with the domain decomposition used in the code, the particle
collisions are automatically efficiently parallelized as long as the particles
are more or less uniformly distributed over the domain.

\subsection{Numerical setup}
\label{subsec:ns}

In our superdroplet simulations,
we consider droplets of radius $10\um$, randomly distributed in space,
together with one droplet of twice the mass, so that the radius is
$2^{1/3}\times 10\um=12.6\um$.
The larger droplet has a higher settling speed than the $10\um$ droplets
and sweeps them up through collision and coalescence.
For each simulation, we track the growth history of the larger droplet
until it reaches $50\um$ in radius and record the time $T$ it takes
to grow to that size.

In the superdroplet algorithm, one usually takes $\nps_i(t_0)\gg1$, which
implies that the actual number of lucky droplets is also more than one.
This was not intended in the original formulation of the lucky
droplet model \citep{Telford1955, Kos05, Wilkinson16} and could allow
the number of superdroplets with heavier (lucky) droplets,
$N_{\rm s}^{\rm(luck)}$, to become larger than unity.
This would manifest itself in the growth history of the lucky droplets
through an increase by more than the mass of a background droplet.
We refer to this as ``jumps''.
Let us therefore now discuss the conditions under which this would
happen and denote the values of $\nps(t_0)$ for the lucky and background
droplets by $\nps_{\rm luck}$ and $\nps_{\rm back}$, respectively.
First, for $\nps_{\rm luck}=\nps_{\rm back}$, the masses of both lucky
and background superdroplets can increase, provided their values
of $\nps(t_0)$ are above unity; see \Fig{fig:2}(c).
Second, even if $\nps_{\rm luck}<\nps_{\rm back}$ initially, new
lucky superdroplets could in principle emerge when the {\em same} two
superdroplets collide with each other multiple times.
This can happen for two reasons.
First, the use of periodic boundary conditions for the superdroplets
(i.e., in the vertical direction in our laminar model with gravity).
Second, two superdroplets can remain at the
same location (corresponding to the same mesh point of the Eulerian grid for the fluid)
during subsequent time steps.
The simulation time step must be less than both the time for a
superdroplet to cross one grid spacing and the mean collision time, i.e., the
inverse collision rate given by \Eq{eq:collision_rate}.
Looking at \Fig{fig:2}, we see that $\nps_{\rm back}$ can then decrease
after each collision and potentially become equal to or drop below the
value of $\nps_{\rm luck}$.
This becomes exceedingly unlikely if initially
$\nps_{\rm back}\gg\nps_{\rm luck}$,
but it is not completely impossible, unless $\nps_{\rm luck}$ is chosen
initially to be unity.

The initial value of $\nps_{\rm back}$ can in principle also be chosen
to be unity.
Although such a case will indeed be considered here, it would defeat
the purpose and computational advantage of the superdroplet algorithm.
Therefore, we also consider the case $\nps_{\rm back}\gg\nps_{\rm luck}$.
As already mentioned, jumps are impossible if $\nps_{\rm luck}$ is unity.
For orientation, we note that the speed of the lucky droplet
prior to the first collision is about $3.5\cm\s^{-1}$,
the average time to the first collision is $490\s$, and thus, it falls
over a distance of about $17\m$ before it collides.

The superdroplet algorithm is usually applied to three-dimensional (3-D) simulations.
If there is no horizontal mixing, one can consider one-dimensional (1-D) simulations.
Moreover, we are only interested in the column in which the lucky droplet resides.
In 3-D, however, the number density of the $10\um$ droplets
beneath the lucky one is in general not the same as the mean number density of
the whole domain.
This leads to yet another element of randomness: fluctuations of
the number density between columns.

Equation~(\ref{dxidt}) is solved with periodic boundary conditions
using the {\sc Pencil Code} \citep{Collaboration2021}, which employs
a third-order Runge-Kutta time stepping scheme.
The superdroplet algorithm is implemented in the {\sc Pencil Code},
which is used to solve \Eqss{eq:pij}{eq:updateN2}.
For the 1-D superdroplet simulations, we employ an initial number density
of background droplets of
$n_0\approx3\times10^8\, {\rm m}^{-3}$ within a
volume $V=L_x\times L_y\times L_z$ with $L_x=L_y=0.002\, \rm{m}$, $L_z=0.214\, \rm{m}$,
and $N_{\rm s}(t_0)=256$ such that the multiplicity is $\xi_{\rm luck}(t_0)=\xi_{\rm back}(t_0)=1$.
For each simulation, 7,686,000 time steps
are integrated with an adaptive time step with a mean value
of $\delta t = 2.942\times 10^{-4}\, \rm{s}$.
For a superdroplet with an initial radius of $12.6\um$ to
grow to $50\um$, 123 collisions are required.
For the purpose of the present study, we designed a parallel technique
to run thousands of 1-D superdroplet simulations simultaneously
\blue{(see details in appendix~\ref{AppPencilApproachI})}.

\section{Lucky-droplet models}
\label{sec:ldm}

\subsection{Basic idea}
\label{BasicIdea}

The LDM describes the collisional growth of a larger droplet that settles
through a quiescent fluid and collides with smaller monodisperse droplets,
that were initially randomly distributed in space.
This corresponds to the setup described in the previous section.
We begin by recalling the main conclusions of \cite{Kos05}.
Initially, the lucky droplet has a radius corresponding to
a volume twice that of the background droplets, whose radius
was assumed to be $r_1=10\um$.
Therefore, its initial radius is $r_2=2^{1/3}r_1=12.6\um$.
After the $(k-1)$th collision step with smaller droplets, it increases as
\begin{equation}
r_k \sim r_1 k^{1/3}.
\label{eq:rk}
\end{equation}
Fluctuations in the length of the time intervals $t_k$ between collision
$k-1$ and $k$ give rise to fluctuating growth histories of the larger droplet.
These fluctuations are quantified by the distribution of the cumulative time
\begin{equation}
T = \sum_{k=2}^{124} t_k,
\label{eq:T}
\end{equation}
corresponding to 123 collisions
needed for the lucky droplet to grow from $12.6\um$ to $50.0\um$
(note that \citet{Kos05} used one more collision, so their final
radius was actually $50.1\um$).
The time intervals $t_k$ between
successive collisions are drawn from an exponential distribution
with a probability $p_k(t_k)=\lambda_k\exp(-\lambda_kt_k)$.
The rates $\lambda_k$ depend on the differential settling velocity
$|\bm{v}_k-\bm{v}_1|$ between the colliding droplets through
\Eqs{eq:pij}{eq:collision_rate}.
Here, however, the background droplets have always the radius $r_1$, so
the collision rate at the $(k-1)$th collision of the lucky droplet
with radius $r_k$ obeys
\begin{equation}
\lambda_k=\pi\left(r_k+r_1\right)^2|\bm{v}_k-\bm{v}_1|
\,E_k\,n,
\label{eq:collision_ratek}
\end{equation}
where $E_k=E(r_k,r_1)$, and $\bm{v}_k$ and $\bm{v}_1$ are approximated by their
terminal velocities.

While the LDM is well suited for addressing theoretical questions
regarding the significance of rare events, it should be emphasized
that it is at the same time highly idealized.
Furthermore, while it is well known that $E_k\ll1$ \citep{Pru78},
it is instructive to assume, as an idealization, $E_k=1$ for
all $k$, so the collision rate \eqref{eq:collision_ratek}
can be approximated as $\lambda_{k}\sim r_k^4$ \citep{Kos05},
which is permissible when $r_k\gg r_1$.
It follows that,
in terms of the collision index $k$, the collision frequency is
\begin{equation}
\lambda_k=\lambda_* k^{4/3},
\label{rk}
\end{equation}
where $\lambda_*=(2\pi/9)(\rho_{\rm d}/{\rho})(gn/\nu) r_1^4$, and $n$ is the
number density of the $10\um$ background droplets.
This is essentially the model of \cite{Kos05} and \cite{Wilkinson16},
except that they also assumed $E_k\ne1$.
They pointed out that, early on, i.e., for small $k$,
$\lambda_k$ is small and therefore the
mean collision time $\lambda_k^{-1}$ is long.
We note that the variance of the mean collision time is $\lambda_k^{-2}$,
which is large for small $k$.
The actual time until the first collision can be very long, but it
can also be very short, depending on fluctuations.
Therefore, at early times, fluctuations have a large impact on the
cumulative collision time.
Note that for droplets with $r\ge 30\um$, the linear Stokes drag
is not valid \citep{Pru78}.

\subsection{Relaxing the power law approximation}
\label{sec:CorrectionsKS05}

\begin{figure}[t!]\begin{center}
\includegraphics[width=\columnwidth]{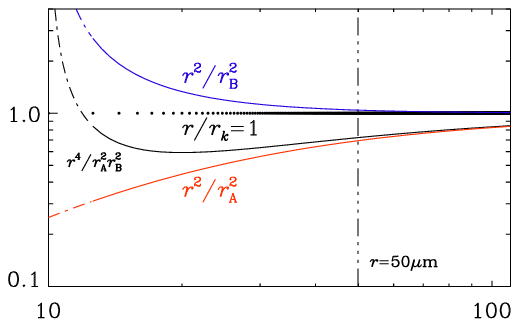}
\end{center}\caption{
Contributions to the two correction factors $r^2/r_{\rm A}^2$ (red)
and $r^2/r_{\rm B}^2$ (blue), as well as their product.
The dashed-dotted parts of the lines apply to radii smaller than $12.6\um$.
The discrete radii $r_k$ for $k\ge2$ are shown in a horizontal line of dots.
The vertical dash-triple-dotted line denote the radius $r=50\um$.
}\label{pt_funct_rArB}
\end{figure}

We now discuss the significance of the various approximations
being employed in the mathematical formulation of the LDM of
\cite{Kos05}.
To relax the approximations made in \Eq{rk}, we now write it in the form
\begin{equation}
\label{eq:mod}
\lambda_k = 
\lambda_* E_k r_{\rm A}^2(r_k) r_{\rm B}^2(r_k)/r_1^4
\quad\mbox{($k\geq2$)},
\end{equation}
where
\begin{equation}
\label{eq:ap1}
r_{\rm A}^2=(r_k+r_1)^2,\quad
r_{\rm B}^2=r_k^2-r_1^2
\end{equation}
would correspond to the expression \Eq{eq:collision_ratek}
used in the superdroplet algorithm.
In \Eq{rk}, however, it was assumed that $r_{\rm A}=r_{\rm B}=r_k$.
To distinguish this approximation from the form used in \Eq{eq:mod},
we denote that case by writing symbolically
``$r_{\rm A}\neq r_k\neq r_{\rm B}$''; see \Fig{pt_funct_rArB}.

In \Eq{eq:ap1}, we have introduced $r_{\rm A}$ and $r_{\rm B}$
to study the effect of relaxing the assumption $r_{\rm A}=r_{\rm B}=r_k$,
made in simplifying implementations of the LDM.
Both of these assumptions are justified at late times when the lucky
droplet has become large compared to the smaller ones, but not early on,
when the size difference is moderate.

\begin{figure}[t!]\begin{center}
\includegraphics[width=\columnwidth]{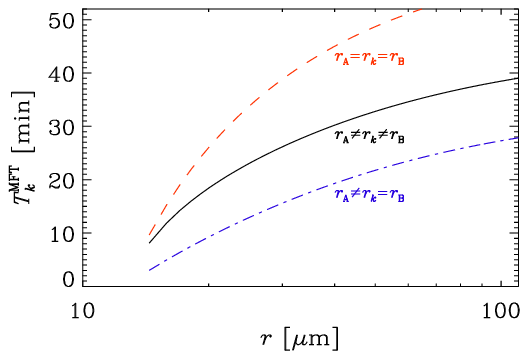}
\end{center}\caption{
Cumulative mean collision times, $T_k^{\rm MFT}$,
for $r_{\rm A}\neq r_k\neq r_{\rm B}$ (solid black line), compared with
the approximations $r_{\rm A}=r_{\rm B}=r_k$ (red dashed line) and only
$r_{\rm B}=r_k$ (blue dash-dotted line).
}\label{pt_funct_MFT}
\end{figure}

By comparison, if fluctuations are ignored, the
collision times are given by $t_k=\lambda_k^{-1}$.
This is what we refer to as mean-field theory (MFT).
In \Fig{pt_funct_MFT} we demonstrate the effect of the contributions from
$r_{\rm A}$ and $r_{\rm B}$ on the mean cumulative collision time in
the corresponding MFT,
\begin{equation}
T_k^{\rm MFT}=\sum_{k'=2}^k t_{k'}^{\rm MFT},
\end{equation}
where
\begin{equation}
t_k^{\rm MFT}=\lambda_k^{-1}
\end{equation}
are the inverse of the mean collision rates.
We see that, while the contribution from $r_{\rm A}$
shortens the mean collision time, that of $r_{\rm B}$
enhances it.
In \Fig{pt_funct_rArB},
we also see that the contributions to the two correction factors
$r^2/r_{\rm A}^2$ and $r^2/r_{\rm B}^2$ have opposite trends,
which leads to partial cancelation in their product.

\begin{figure}[t]
\begin{center}
\includegraphics[width=0.5\textwidth]{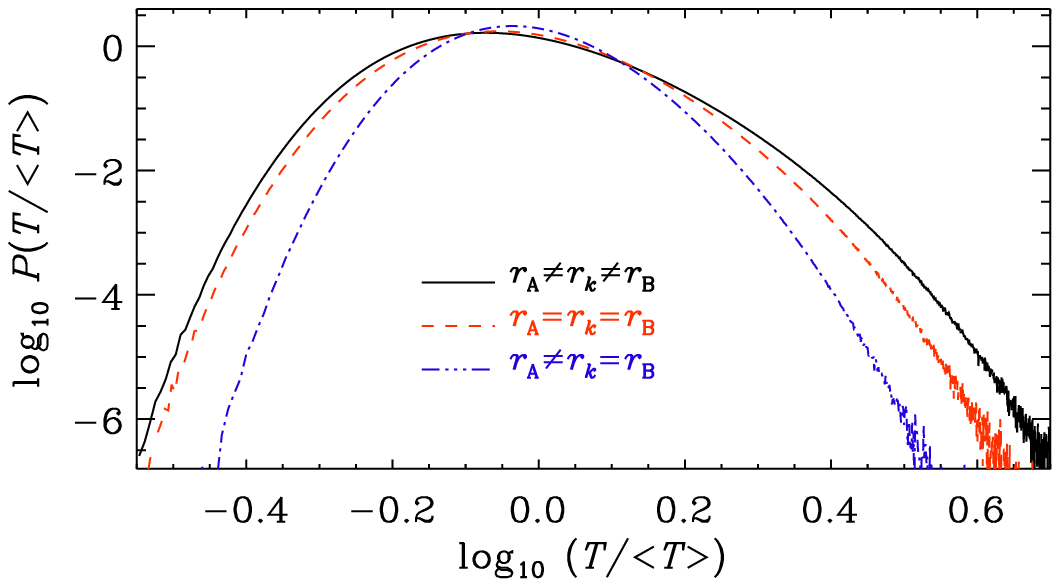}
\end{center}
\caption{\label{fig:pcomp_mom}
Comparison of $P(T)$ in a double-logarithmic representation for the LDM
appropriate to our benchmark (black solid line) with various approximations
where $r_{\rm A}=r_{\rm B}=r_k$ (red dashed line) along with a case where only
$r_{\rm B}=r_k$ is assumed (blue dash-dotted line).
Here we used $10^{10}$ realizations.
Note that we plot the distribution of the cumulative times
versus the normalized time, $T/\bra{T}$, as was done in the work
of \cite{Kos05}.
Normalizing by $\bra{T}$ allows us to see changes in the shape of
$P(T/\bra{T})$, thus allows a more direct comparison of the subtle
differences in the shapes of the different curves and ensures that
the peaks of all curves are at approximately the same position.
}
\end{figure}

In \Fig{fig:pcomp_mom} we show a comparison of the distribution of
cumulative collision times for various representations of $r_k$.
Those are computed numerically using $10^{10}$ realizations of sequences
of random collision times $t_k$.
We refer to appendix~\ref{AppPencilApproachI} for details of
performing this many realizations.

\begin{figure}[t]
\begin{center}
\includegraphics[width=0.5\textwidth]{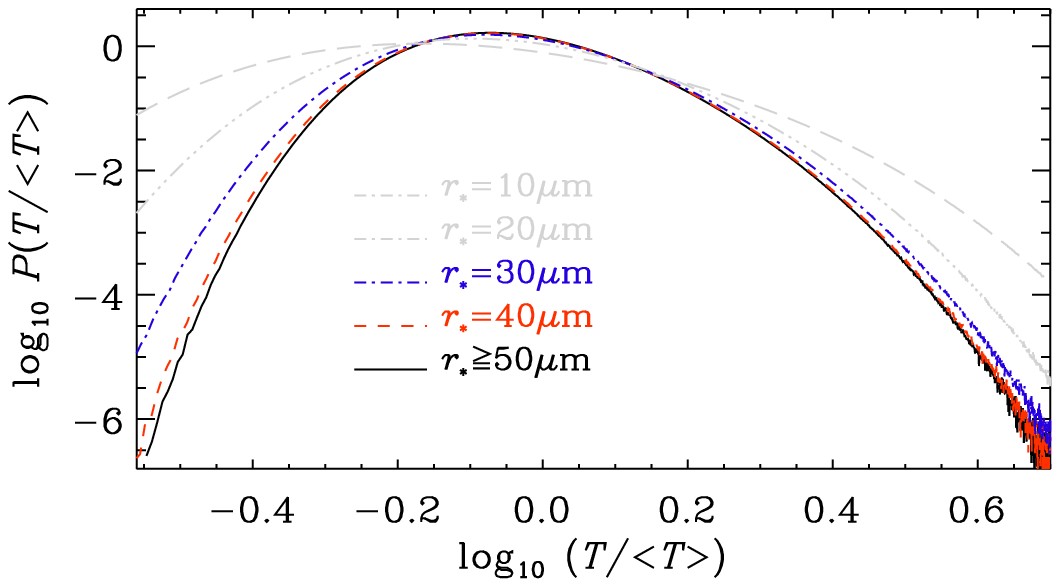}
\end{center}
\caption{\label{fig:pcomp_mom_effi}
Comparison of $P(T)$ in a double-logarithmic representation for the LDM
for $r_*=40\um$ and $30\um$ using $r_{\rm A}\neq r_k\neq r_{\rm B}$.
The black line agrees with that in \Fig{fig:pcomp_mom}, and the two
gray lines refer to the cases with $r_*=20\um$ and $10\um$.
Here we used $10^{10}$ realizations.
}
\end{figure}

The physically correct model is where $r_{\rm A}\neq r_k\neq r_{\rm B}$
(black line in \Fig{fig:pcomp_mom}).
To demonstrate the sensitivity of $P(T)$ to changes in the representation of $r_k$,
we show the result for the approximations $r_{\rm A}=r_k=r_{\rm B}$
(red line) and $r_{\rm A}\neq r_k=r_{\rm B}$ (blue line).
The $P(T)$ curve is also sensitive to changes in the collision efficiency
late in the evolution.
To demonstrate this, we assume $E_k\propto r_k^2$ when $r_k$
exceeds a certain arbitrarily chosen value $r_*$ between $10$ and $40\um$,
and $E_k={\rm const}$ below $r_*$ \citep{lamb_verlinde_2011}.
To ensure that $E_k\leq1$, we take
\begin{equation}
E_k=E_*\,\max\left(1,\,(r_k/r_*)^2\right),
\end{equation}
with $E_*=(r_*/50\um)^2$.
However, the normalized $P(T)$ curves are independent of the
choice of the value of $E_*$.
In \Fig{fig:pcomp_mom_effi}, we show the results for
$r_{\rm A}\neq r_k\neq r_{\rm B}$ using $r_*=40\um$ and $30\um$
(red and blue lines, respectively) and compare with the case
$E_k={\rm const}$.
The more extreme cases with $r_*=20\um$ and $10\um$ are shown
as gray lines.
The latter is similar to the case $\lambda_{k}\sim r_k^6$
considered by \cite{Kos05} and \cite{Wilkinson16}.

\begin{table}[t!]
\caption{
Moments of $X=\ln(T/\bra{T})$ computed from $10^{10}$ realizations for
different values of $r_\ast$ (in $\mu$m), and different prescriptions
of $r_{\rm A}$ and $r_{\rm B}$.
The corresponding values of $T_{123}^{\rm MFT}$ are also given
and are normalized to unity for $r_{\rm A}\neq r_k\neq r_{\rm B}$
with $r_*\ge50\um$.
}\centering
\begin{tabular}{cccccccr}
\hline\hline
$r_\ast$ & $r_{\rm A}$ & $r_{\rm B}$ & $T_{123}^{\rm MFT}$ &
$\bra{X}$ & $\!\sigma(X)\!$ & $\!\!\!\SKEW X\!\!\!$ & $\!\!\KURT X\!\!$ \\
\hline
---&  ---  & $r_k$ & 0.67 & $-0.020$ & 0.21 &\!\! 0.22 \!\!&\!\! $ 0.08$ \\
---& $r_k$ & $r_k$ & 1.49 & $-0.033$ & 0.25 &\!\! 0.25 \!\!&\!\! $ 0.05$ \\
---&  ---  &  ---  &  1   & $-0.040$ & 0.28 &\!\! 0.34 \!\!&\!\! $ 0.10$ \\
40 &  ---  &  ---  & 0.99 & $-0.041$ & 0.28 &\!\! 0.33 \!\!&\!\! $ 0.09$ \\
30 &  ---  &  ---  & 0.93 & $-0.046$ & 0.30 &\!\! 0.28 \!\!&\!\! $ 0.05$ \\
20 &  ---  &  ---  & 0.79 & $-0.063$ & 0.35 &\!\! 0.18 \!\!&\!\! $-0.04$ \\
10 &  ---  &  ---  & 0.34 & $-0.111$ & 0.47 &\!\! 0.16 \!\!&\!\! $-0.17$ \\
\hline\hline
\end{tabular}
\label{tab:Moments}
\end{table}

When $r_{\rm A}=r_k=r_{\rm B}$, or only $r_k=r_{\rm B}$, the $P(T)$
curves exhibit smaller widths.
By contrast, when the collision efficiency becomes quadratic later
on (when $r>r_*\equiv30\um$ or $40\um$), the $P(T)$
curves have larger widths; see \Fig{fig:pcomp_mom_effi}.
To quantify the shape of $P(T)$, we give in Table~\ref{tab:Moments}
the average of $X\equiv\ln(T/\bra{T})$, its standard deviation
$\sigma=\bra{x^2}^{1/2}$, where $x\equiv X-\bra{X}$,
its skewness $\SKEW X=\bra{x^3}/\sigma^3$, and
its kurtosis $\KURT X=\bra{x^4}/\sigma^4-3$.
We recall that, for a perfectly lognormal distribution,
$\SKEW X=\KURT X=0$.
The largest departure from zero is seen in the skewness, which is
positive, indicating that the distribution broadens for large $T$.
The kurtosis is rather small, however.

The main conclusion that can be drawn form the investigation mentioned above
is that, as far as the shapes of the different curves are concerned,
it does not result in any significant error to assume $r_k\gg r_1$.
The value of $\sigma$ is only about 10\% smaller if
$r_{\rm A}=r_k=r_{\rm B}$ is used (compare the red dashed
and black solid lines in \Fig{fig:pcomp_mom}).
This is because the two inaccuracies introduced by $r_{\rm A}$ and
$r_{\rm B}$ almost cancel each other.
When $r_*=40\um$ or $30\um$, for example, the values of $\sigma$
increase by 3\% and 15\%, respectively; see Table~\ref{tab:approach},
where we also list the corresponding values of $T_{124}^{\rm MFT}$.
On the other hand, the actual averages such as
$\bra{T}\approx T_{124}^{\rm MFT}$ vary by almost 50\%.

A straightforward extension of the LDM is to take
horizontal variations in the local column density into account.
Those are always present for any random initial conditions, but could
be larger for turbulent systems, regardless of the droplet speeds.
In 3-D superdroplet simulations, large
droplets can fall in different vertical columns that contain different
numbers of small droplets, a consequence of the fact that the small
droplets are initially randomly distributed.
To quantify the effect of varying droplet number densities in
space, it is necessary to solve
for an ensemble of columns with different number densities of the
$10\um$ background droplets and compute the distribution of
cumulative collision times.
These variations lead to a broadening of $P(T)$, but it is a priori
not evident how important this effect is.
A quantitative analysis is given in appendix~\ref{app:3dLDM}.

\subsection{Relation to the superdroplet algorithm}

To understand the nature of the superdroplet algorithm, and why it captures
the lucky droplet problem accurately, it is important to realize that
the superdroplet algorithm is actually a combination of two separate
approaches to solving the LDM, each of which turns out to be able
to reproduce the lucky droplet problem to high precision.
In principle, we can distinguish four 
different approaches (\Tab{tab:approach}) to obtaining the collision time interval $t_k$.
In approach~I, $t_k$ was taken from an exponential distribution of random numbers.
Another approach is to use a randomly distributed set of $10\um$
background droplets in space and then determine the distance to the
next droplet within a vertical cylinder of possible collision partners
to find the collision time (approach~II).
A third approach is to use the mean collision rate to compute
the probability of a collision within a fixed time interval.
We then use a random number between zero and one
\citep[referred to as Monte Carlo method; see, e.g.,][]{Sok97}
to decide whether at any time there is a collision or not (approach~III).
This is actually what is done within each grid cell in the superdroplet algorithm; see
\Eqs{eq:pij}{eq:collision_rate}.
The fourth approach is the superdroplet algorithm discussed extensively
in section~\ref{sec:method}.\ref{sec:alg} (approach~IV).
It is essentially a combination of approaches~II and III.
We have compared all four approaches and found that they all give very
similar results.
In the following, we describe approaches~II and III in more detail,
before focussing on approach~IV in section~\ref{sec:gh}.

\begin{table}[t!]
\caption{
Summary of the four approaches.
}\centering
\begin{tabular}{ll}
\hline\hline
Approach \!\! & Description\\
\hline
I   &  time interval $t_k$ drawn from distribution \\
II  &  primitive Lagrangian particles collide \\
III &  probabilistic, just a pair of superdroplets \\
IV  &  superdroplet model (combination of II \& III)\\
\hline\hline
\end{tabular}
\label{tab:approach}
\end{table}

\subsection{Solving for the collisions explicitly}

A more realistic method (approach~II; see \Tab{tab:approach}) is to compute
random realizations of droplet positions in a tall box of
size $L_{\rm h}^2\times L_z$, where $L_{\rm h}$ and $L_z$ are the
horizontal and vertical extents, respectively.
We position the lucky droplet in the middle of the top plane of the box.
Collisions are only possible within a vertical cylinder of radius $r_k+r_1$
below the lucky droplet.
Next, we calculate the distance $\Delta z$ to the first collision partner
within the cylinder.
We assume that both droplets reach their terminal velocity well
before the collision.
This is an excellent approximation for dilute systems such as clouds,
because the droplet response time $\tau_k$ of \Eq{response_time}
is much shorter than the mean collision time.
Here we use the subscript $k$ to represent the time until
the $(k-1)$th collision, which is equivalent to the $i$th droplet.
We can then assume the relative velocity between
the two as given by the difference of their terminal velocities as
\begin{equation}
  \Delta v_k=(\tau_k-\tau_1)\,g.
\label{terminal}
\end{equation}
The time until the first collision is then given by $t_2=\Delta z/\Delta v_2$.
This collision results in the lucky droplet having increased its
volume by that of the $10\um$ droplet.
Correspondingly, the radius of the vertical cylinder of collision
partners is also increased.
We then search for the next collision partner beneath the position of
the first collision, using still the original realization of $10\um$ droplets.
We continue this procedure until the lucky droplet reaches a radius of $50\um$.
\blue{Approach II is an explicit method compared to other approaches listed in
\Tab{tab:approach}.}

\subsection{The Monte Carlo method to compute $t_k$}
\label{MonteCarloMethod}

In the Monte Carlo method (approach III; see \Tab{tab:approach}) we choose a time step $\delta t$
and step forward in time.
As in the superdroplet algorithm, the probability of a collision
is given by $p_k=\lambda_k\delta t$; see \Eq{eq:pij}.
We continue until a radius of $50\um$ is reached.
We note that in this approach, $n$ is kept constant, i.e., no background
droplet is being removed after a collision.

Approach~III also allows us to study the effects of jumps in the droplet size
by allowing for several lucky droplets at the same time
and specifying their collision probability appropriately.
These will then be able to interact not only with the $10\um$ background
droplets, but they can also collide among themselves, which causes the
jumps.
We will include this effect in solutions of the LDM using approach~III
and compare with the results of the superdroplet algorithm.

\section{Results}
\label{sec:gh}

\subsection{Accuracy of the superdroplet algorithm}
\label{sec:Accuracy}

We now want to determine to what extent the fluctuations are
correctly represented by the superdroplet algorithm.
For this purpose, we now demonstrate the degree of quantitative
agreement between approaches~I--III and the corresponding
solution with the superdroplet algorithm (approach~IV; see \Tab{tab:approach}).
This is done by tracking the growth history of each lucky droplet.
As the first few collisions determine the course of the formation
of larger droplets, we also use the distribution $P(T)$
of cumulative collision times $T$.
We perform $N_{\rm real}$ superdroplet simulations with different
random seeds using $\nps_i(t_0)=1$.

\begin{figure}[t]\begin{center}
\includegraphics[width=0.48\textwidth]{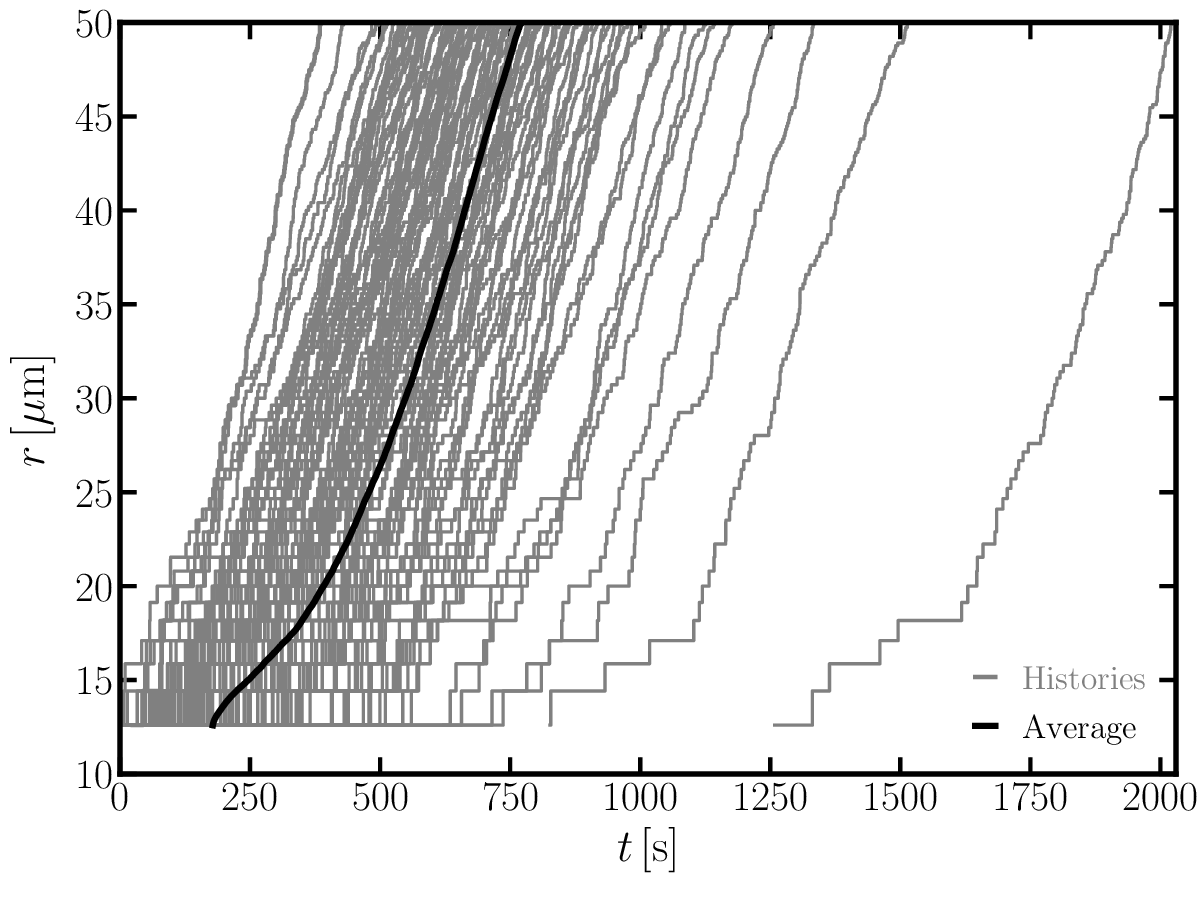}
\end{center}\caption{98 growth histories of lucky droplets
obtained from 98 independent 1-D superdroplet simulations
(approach~IV), as described in the text.
All superdroplets have initially the same number of droplets,
$\nps_i(t_0)=1$ with $N_{\rm s}(t_0)=256$. The mean number density of
droplets is $n_0=3\times10^8\,\rm{m}^{-3}$.
The thick solid line shows the average time for each radius.
}\label{pplucky_select_50history_assemble_1D_lucky_data_191206}
\end{figure}

\begin{figure}[t]\begin{center}
\includegraphics[width=0.5\textwidth]{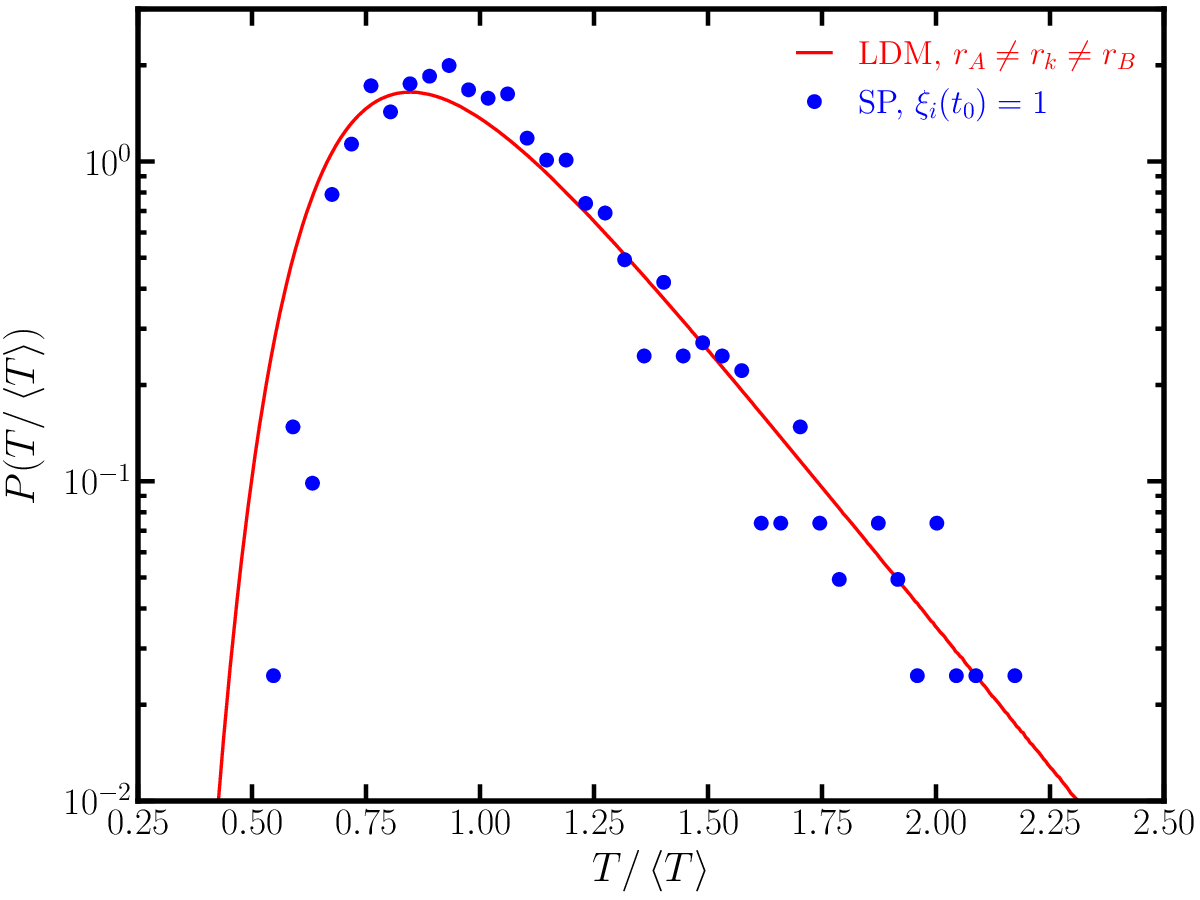}
\end{center}\caption{
Corresponding $P(T)$ of \Fig{pplucky_select_50history_assemble_1D_lucky_data_191206}
obtained with the superdroplet algorithm (blue dots) and
the LDM using approach~I with $r_{\rm A}\neq r_k\neq r_{\rm B}$ (red solid line).
}\label{pT50mum_SP_theory_1D}
\end{figure}

\begin{figure}[t]\begin{center}
\includegraphics[width=0.5\textwidth]{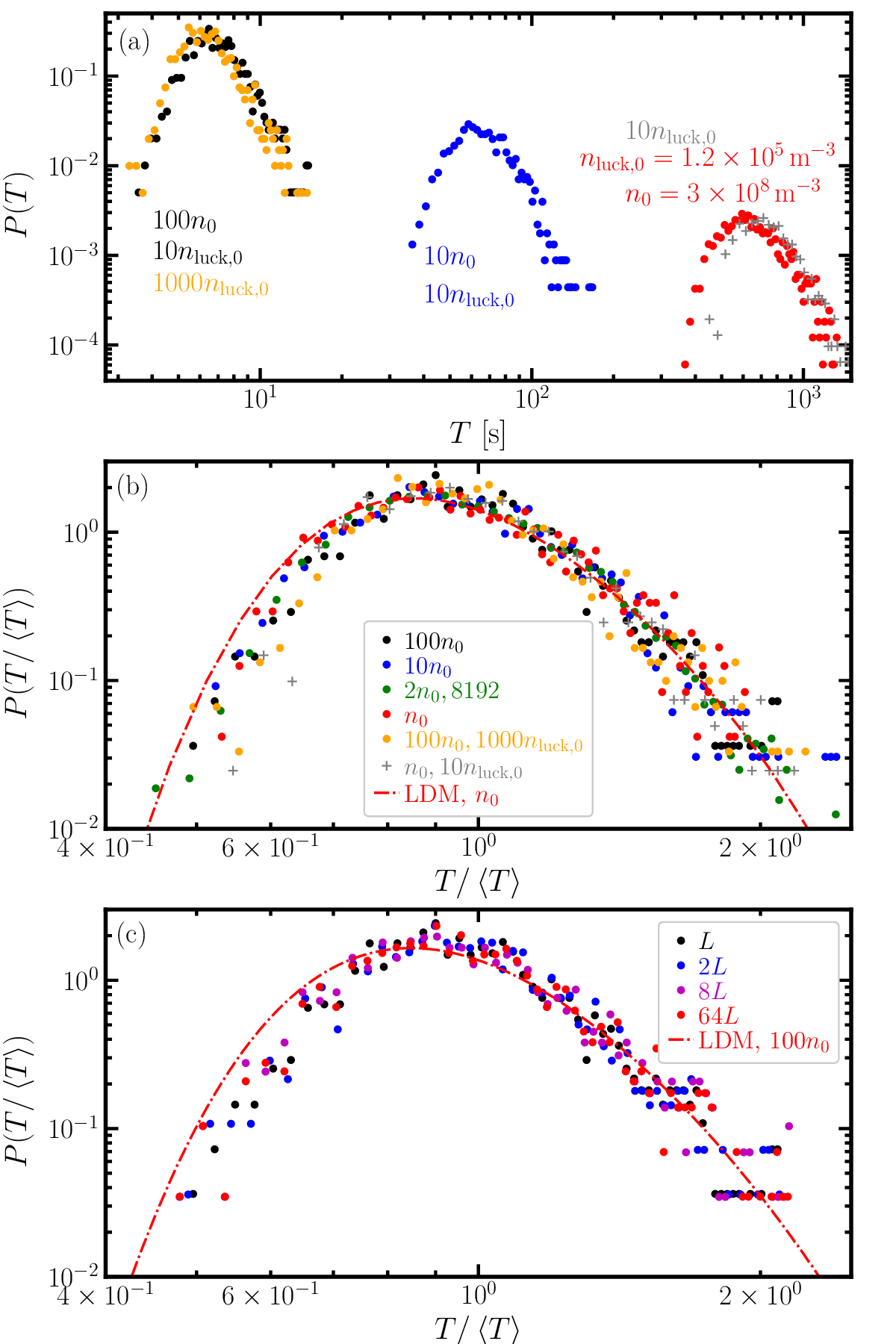}
\end{center}\caption{
(a): $P(T)$ for $n_0$ (red), $10\,n_0$ (blue), and $100\,n_0$ (black) with
$n_0=3\times10^8\, \rm{m}^{-3}$ and $L=0.214\, \rm{m}$.
In the last case, 0.5\% of the background droplets were removed;
the orange symbols denote a case with 100 times larger value of $n_{\rm luck}$,
where 50\% of the background droplets were removed.
\blue{
The gray symbols denote the case of \Fig{pT50mum_SP_theory_1D} with
ten times more physical lucky droplets ($10\,n_{\rm luck,0}$).
}
(b): $P(T/\bra{T})$ for $n=n_0$ (red), $2\,n_0$ (green), $10\,n_0$ (blue),
$100\,n_0$ with $10\,n_{\rm luck}$ (black), $1000\,n_{\rm luck}$ (orange),
\blue{
and $n_0$ with $10\,n_{\rm luck,0}$ (gray).
}
(c): $P(T/\bra{T})$ for $L$, $2L$, $8L$, and $64L$ with $100\,n_0$,
obtained using the superdroplet algorithm (approach~IV).
The red dash-dotted line in (b) represents the LDM (approach~I) with
$r_{\rm A}\neq r_k\neq r_{\rm B}$ and $n_0=3\times10^8\, \rm{m}^{-3}$,
which is the same simulation as the one in \Fig{pT50mum_SP_theory_1D}.
The green dots in (b) is for 8192 realizations,
while all the other simulations are for 1024 realizations.
}\label{preached_comp}
\end{figure}

We begin by looking at growth histories for many individual realizations
obtained from the superdroplet simulation.
\Fig{pplucky_select_50history_assemble_1D_lucky_data_191206}
shows an ensemble of growth histories (thin gray lines)
obtained from $N_{\rm real}\approx10^3$ independent simulations, as described above.
The times between collisions are random, leading to a
distribution of cumulative growth times to reach $50\,\mu{\rm m}$.
Also shown is the mean growth curve (thick black line), obtained
by averaging the time at fixed radii $r$.
This figure demonstrates that the fluctuations are substantial.
We also see that large fluctuations relative to the average time are rare.

To quantify the effect of fluctuations from all realizations, we now consider
the corresponding $P(T)$ in \Fig{pT50mum_SP_theory_1D}.
We recall that $\nps_i(t_0)=1$ for our superdroplet simulation in
\Fig{pT50mum_SP_theory_1D}. However, a simulation with $\nps_i(t_0)=50$
yields almost the same result; see appendix~\ref{app:nds}.

The comparison of the results for the LDM using approach~I and the
superdroplet algorithm shows small differences.
The width of the $P(T)$ curve is slightly larger for approach~I than
for the superdroplet simulations.
This suggests that the fluctuations, which are at the heart of the
LDM, are slightly underrepresented in the superdroplet algorithm.
However, this shortcoming may also be a consequence of our choice
of having used only 256 superdroplets, i.e., one lucky and 255 background superdroplets.
Given that the multiplicities of lucky and background droplets was unity,
each collision removed one background droplet.
Thus, after 123 collisions, almost $50\%$ of the background droplets
were removed by the time the lucky droplet reached $50\um$.
Nevertheless, as we will see below, this has only a small effect.





\begin{table}[b]
\begin{center}  
\caption{\label{tab:color}
\blue{
Runs of Fig.~8(a), where
$n_i$ and $n_{i,\rm luck}$ are in units of $n_{i0}=7.5\times10^7\m^{-3}$,
$n$ is in units of $n_0=3\times10^8\m^{-3}$,
$n_{\rm luck,0}$ is in units of $1.2\times10^5\m^{-3}$,
using $\delta x^3=2.6\times10^{-8}\m^3$ for all runs,
except for the last one (gray), where it is a factor 2 smaller.
}
}\begin{tabular}{lccccc}
\hline\hline
color in Fig.~8(a)                          & red & blue & black & orange &gray\\
\hline\hline
$n_i/n_{i0}$                                & $1$ & $10$ & $100$ &  $100$ &  1 \\
$n_{i,\rm luck}/n_{i0}$                     & 0.1 &   1  &   1   &   100  &  1 \\
$n/n_0$                                     & $1$ & $10$ & $100$ &  $100$ &  1 \\
$n_{\rm luck}/n_{\rm luck,0}$               & $1$ & $10$ &  $10$ & $1000$ & 10 \\
$\xi_{\rm back}=\delta x^3\,n_i$            &  2  &  20  & 200   &  200   &  1 \\
$\xi_{\rm luck}=\delta x^3\,n_{i,\rm luck}$ & 0.2 &   2  &   2   &  200   &  1 \\
removed fraction                            & $5\%$ & $5\%$ & $0.5\%$ &  $50\%$   & $50\%$\\
\hline\hline
\end{tabular}
\end{center}
\end{table}

An important question is to what extent our results depend on the
number density of background droplets and the size of the computational domain.
To examine this with the superdroplet algorithm (approach~IV),
we consider three values of the initial number density: $n_0=3\times10^8\m^{-3}$,
$10\,n_0$, and $100\,n_0$, while the initial number density of the lucky
droplet is $n_{\rm luck,0}\equiv1.2\times10^5\m^{-3}$, $10\,n_{\rm luck,0}$, and
again $10\,n_{\rm luck,0}$, respectively; \blue{see \Tab{tab:color} for a summary.}
Thus, even though the lucky droplet has to collide 123 times to reach $50\um$,
it only removes $123\,n_{\rm luck}/n_0=5\%$, 5\%, and 0.5\% of the droplets, respectively.
\Fig{preached_comp} shows $P(T)$ for these
three cases using first the cumulative time $T$ [\Fig{preached_comp}(a)]
and then the normalized time $T/\bra{T}$ [\Fig{preached_comp}(b)].
We see that the positions of the peaks in $P(T)$ change linearly
with the initial number density $n_0$, but $P(T/\bra{T})$
are very similar to each other.
This is related to the fact that, after normalization, $n_0$ drops out
from the expression for $t_k/\bra{T}$ in the LDM (approach~I); see \Eq{eq:T}.
At small values of $T/\bra{T}$, however, all curves show a similar
slight underrepresentation of the fluctuations as already seen in
\Fig{pT50mum_SP_theory_1D}.
In all these simulations, we used 1024 realizations, except \blue{possibly} for
one case where we used 8192 realizations; see the green symbols
in \Fig{preached_comp}(b).
The distribution of cumulative growth times is obviously much smoother
in the latter case, but the overall shape is rather similar.

In the above, the number density of the lucky droplets has been much
smaller than the number density of the background droplets.
This means that for each collision the physical number of background
droplets changed by only a small amount (5\% or 0.5\%).
To see how sensitive our results for $P(T)$ are to this number, we now
perform an extra experiment where $50\%$ of the background droplets
are removed by the time the lucky droplet reaches $50\um$.
This is also shown in \Fig{preached_comp}(a) and (b);
see the orange symbols\blue{, where $\xi_{\rm luck}=\xi_{\rm back}=200$}.
We see that even for 50\% removal the results are essentially unchanged.

In our superdroplet simulations (approach~IV; see \Tab{tab:approach}), the vertical extent of the
simulation domain is only $L=0.214\m$.
This is permissible given that we use periodic boundary conditions
for the particles.
Nevertheless, the accuracy of our results may suffer from poor statistics.
To investigate this in more detail, we now perform
1-D simulations with $2L$, $8L$, and $64L$.
At the same time, we increased the number of mesh points and the
number of superdroplets by the same factors.
Since the shape of $P(T/\bar{T})$ is almost independent of
$n_0$, as shown in \Fig{preached_comp}(b),
we use $n_0=3\times10^{10}\m^{-3}$ instead of $n_0=3\times10^8\m^{-3}$
to reduce the computational cost.
As shown in \Fig{preached_comp}(c), $P(T/\bar{T})$ is insensitive to the domain
size.
Therefore, our results with $L=0.214\m$ can be considered
as accurate with respect to $P(T/\bar{T})$.

In the following, we discuss how our conclusions relate to those of
earlier work.
We then discuss a number of additional factors that can modify the results.
Those additional factors can also be taken into account in the LDM.
Even in those cases, it turns out that the differences between the LDM
and the superdroplet algorithm are small.

\begin{figure}[t]
\begin{center}
\includegraphics[width=0.5\textwidth]{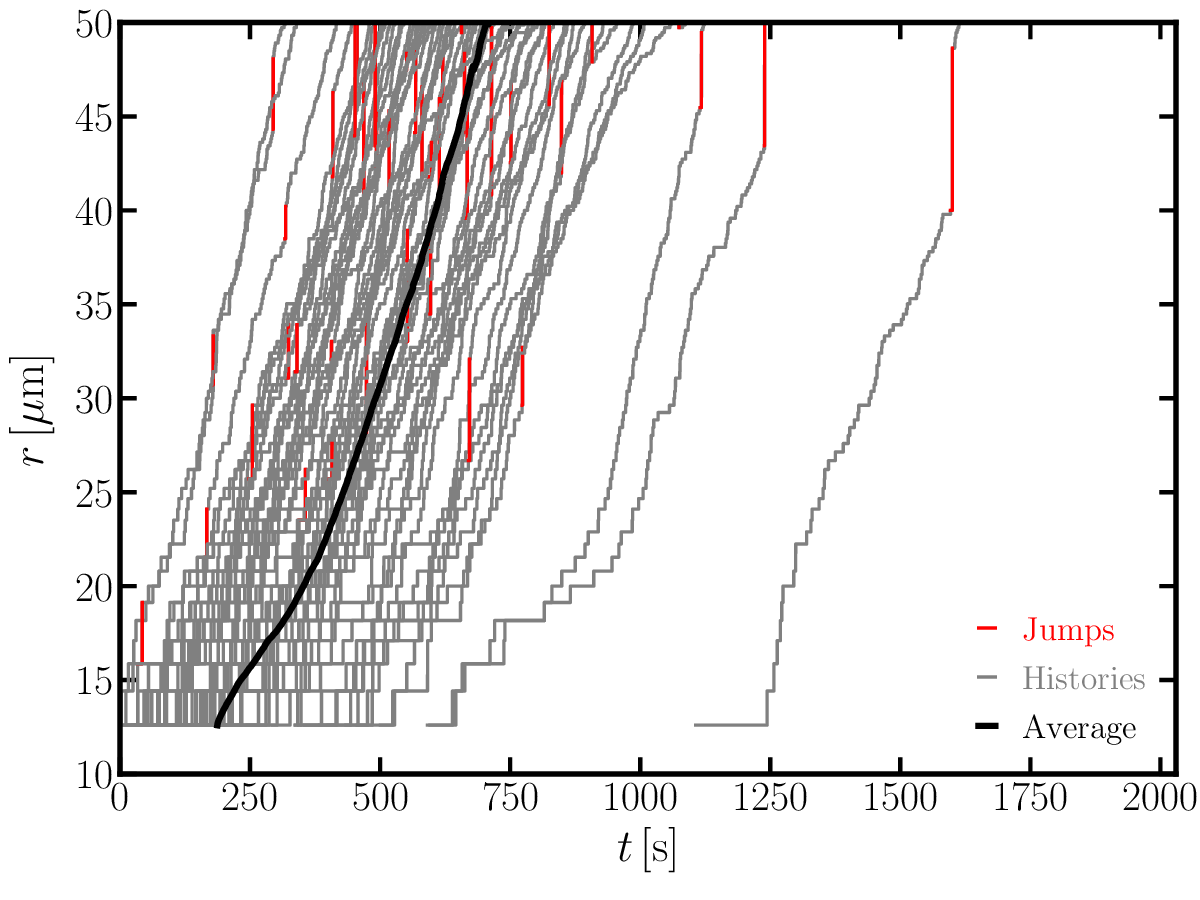}
\end{center}
\caption{\label{fig:ensemble} Same as
\Fig{pplucky_select_50history_assemble_1D_lucky_data_191206}
but with initial condition $\nps_i(t_0)=2$ using $N_{\rm s}(t_0)=128$,
corresponding to the same number of physical droplets as in
\Fig{pplucky_select_50history_assemble_1D_lucky_data_191206},
where $\nps_i(t_0)=1$.
Note the occurrence of jumps, indicated in red.}
\label{pplucky_50history_mean_lucky_poisson_t_Ns128_N2p5e2_lucky12p6_grav_1D_NR}
\end{figure}

\subsection{The occurrence of jumps}
\label{occ_jumps}

One of the pronounced features in our superdroplet simulations with
$\nps_i(t_0)>1$ is the possibility of jumps.
We see examples in \Fig{fig:ensemble} where $\nps_{\rm luck}=\nps_{\rm back}=2$ and the
jumps are visualized by the red vertical lines.
Those jumps are caused by the coagulation of the lucky droplet with droplets of radii
larger than $10\um$ that were the result of other lucky droplets in
the simulations.
What is the effect of these jumps?
Could they be responsible for the behavior found by \citet{Dziekan17}
that the difference in their $t_{10\%}$ between the numerical
and theoretical calculation decreases with the square root of the number
of physical droplets, as we discussed in section~\ref{Introduction}?

\begin{figure}[t]
\begin{center}
\includegraphics[width=.5\textwidth]{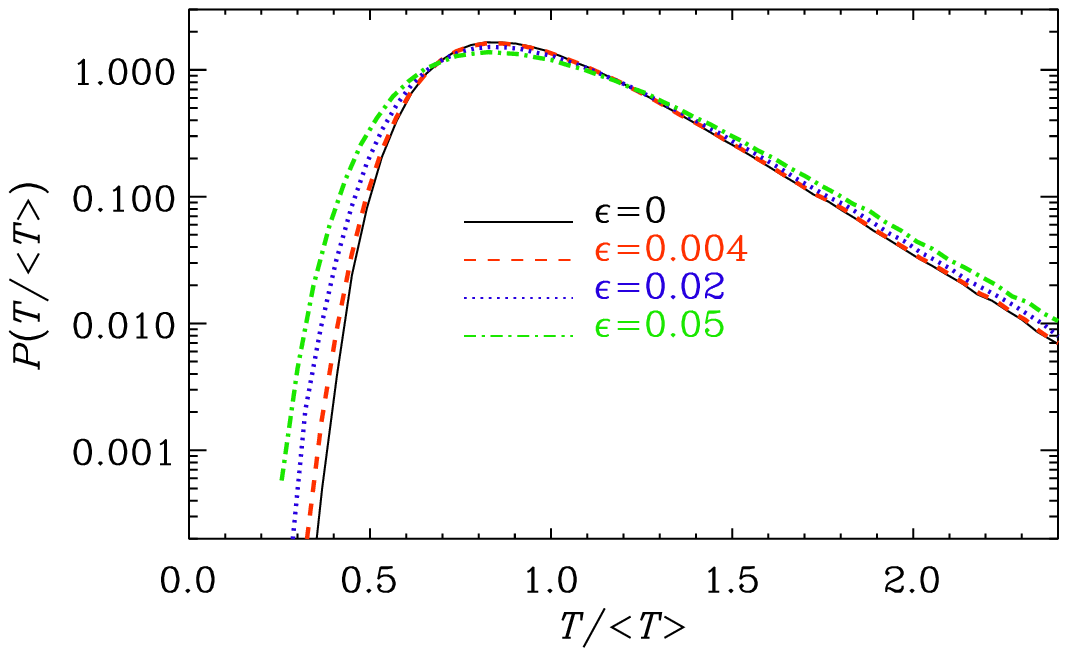}
\end{center}
\caption{\label{fig:pcomp_dens_all_mult_b}
Comparison of models with $\epsilon=0$ (no jumps),
0.004 (the value expected for the simulations), 0.02, and 0.05
using approach~III; see \Tab{tab:approach}.
}
\end{figure}

It is clear that those jumps occur mainly during
the last few steps of a lucky droplet growing to $50\um$
(see \Fig{pplucky_50history_mean_lucky_poisson_t_Ns128_N2p5e2_lucky12p6_grav_1D_NR})
when there has been enough time to grow several more lucky droplets.
Because the collision times are so short at late times, the jumps are
expected to be almost insignificant.
To quantify this, it is convenient to use approach~III, where we
choose $N_{\rm s}^{\rm(luck)}=3$ superdroplets simultaneously.
(As always in approach~III, the background particles are still represented
by only one superdroplet, and $n$ is kept constant.)
We also choose $\nps_{\rm luck}=1$, and therefore $N_{\rm d}^{\rm(luck)}=3$.
The lucky droplets can grow through collisions with the $10\um$ background droplets
and through mutual collisions between lucky droplets.
The collision rate between lucky droplets $i$ and $j$ is,
analogously to \Eq{eq:mod}, given by
\begin{equation}
\lambda_{ij}^{\rm(luck)}=\pi\left(r_i+r_j\right)^2|\vec{v}_i-\vec{v}_j|
\; n_{\rm luck},
\label{eq:collision_rate_luck}
\end{equation}
where $n_{\rm luck}$
is the number density of physical droplets in
the superdroplet representing the lucky droplet.
To obtain an expression for $n_{\rm luck}$ in terms of the
volume of a grid cell $\delta x^3$,
we write $n_{\rm luck}=\nps_{\rm luck}/\delta x^3$.
The ratio of the
physical number of lucky droplets, $N_{\rm d}^{\rm(luck)}$, to the
physical number of background droplets, $N_{\rm d}^{\rm(back)}$
is given by
\begin{equation}
\epsilon=\frac{N_{\rm d}^{\rm(luck)}}{N_{\rm d}^{\rm(back)}}
=\frac{\nps_{\rm luck}\,N_{\rm s}^{\rm(luck)}}{\nps_{\rm back}\,N_{\rm s}^{\rm(back)}}.
\label{ratio}
\end{equation}
To investigate the effect of jumps on $P(T)$ in the full superdroplet
model studied above (see
\Figs{pplucky_select_50history_assemble_1D_lucky_data_191206}
{pplucky_50history_mean_lucky_poisson_t_Ns128_N2p5e2_lucky12p6_grav_1D_NR}),
we first consider the case depicted in
\Fig{pplucky_select_50history_assemble_1D_lucky_data_191206},
where $\nps_{\rm luck}=\nps_{\rm back}\equiv\nps_i(t_0)=1$.
Here, we used $N_{\rm s}=256$ superdroplets, of which one contained the lucky droplet,
so $N_{\rm s}^{\rm(luck)}=1$, and the other
255 superdroplets contained a $10\um$ background droplet each.
In our superdroplet solution, the ratio (\ref{ratio}) was therefore
$\epsilon\approx1/255=0.004$.
Using approach~III, $\epsilon$ enters simply as an extra factor
in the collision probability between different lucky droplets.
(In approach III, all quantities in \Eq{ratio} are kept constant.)
The effect on $P(T)$ is shown in \Fig{fig:pcomp_dens_all_mult_b}, where we present
the cumulative collision times for models with three values of $\epsilon$
using approach~III.
We see that for small values of $\epsilon$, the
cumulative distribution function is independent of $\epsilon$,
and the effect of jumps is therefore negligible
(compare the black solid and the red dashed lines of \Fig{fig:pcomp_dens_all_mult_b}).
More significant departures due to jumps can be seen
when $\epsilon=0.02$ and larger.

\begin{figure}[t]
\begin{center}
\includegraphics[width=0.5\textwidth]{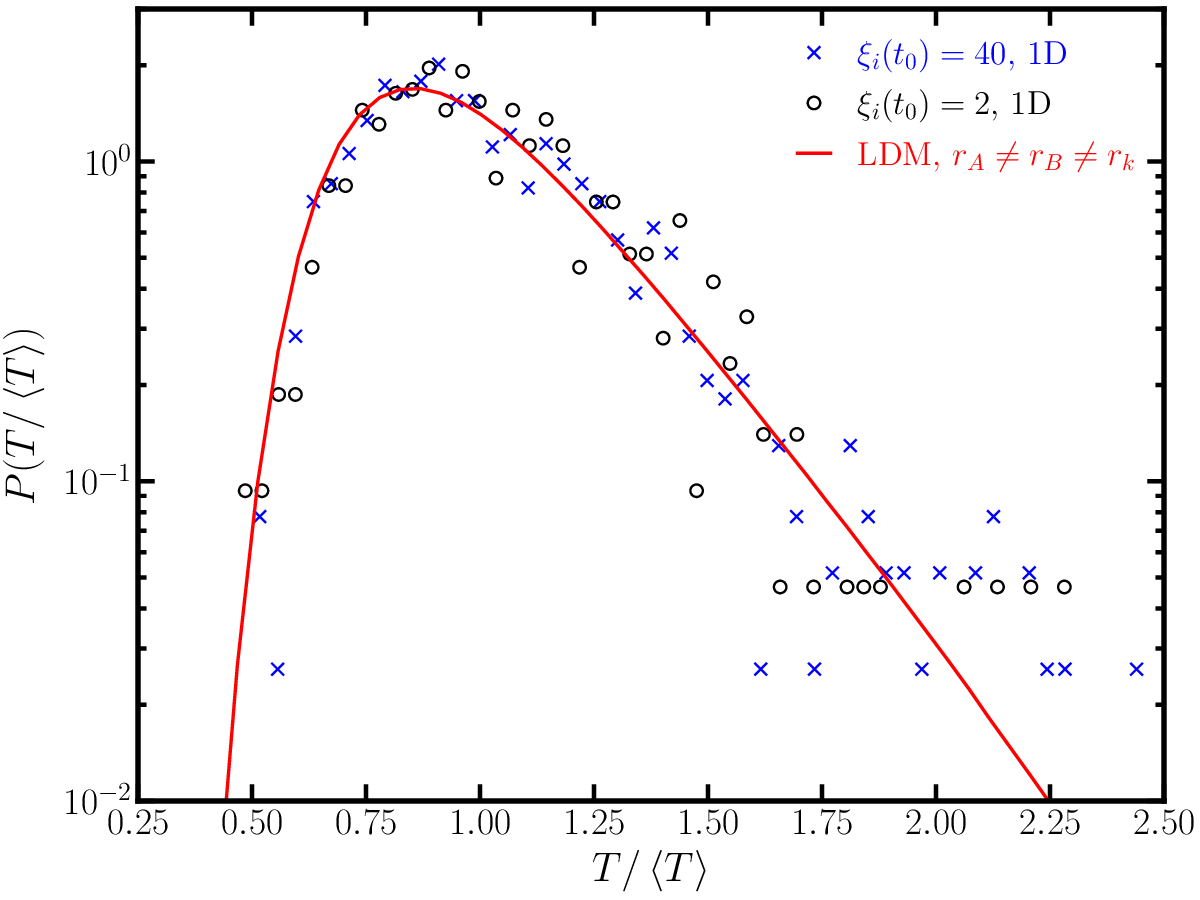}
\end{center}
\caption{
$P(T/\langle T\rangle)$ of simulations in
\Fig{pplucky_50history_mean_lucky_poisson_t_Ns128_N2p5e2_lucky12p6_grav_1D_NR} (black circles)
and the ones with initially $\nps_{\rm back}=40$ (blue crosses).
$\nps_{\rm luck}=2$ in both cases.
The red line denotes the LDM (approach~I) with $r_{\rm A}\neq r_k\neq r_{\rm B}$,
which is the same simulation as the one in \Fig{pT50mum_SP_theory_1D}.
}
\label{fig:pT50mum_SP_theory_1D_Nps}
\end{figure}

Let us now compare with the case in which we found
jumps using the full superdroplet approach (approach~IV).
The jumps in the growth histories cause the droplets to grow faster
than without jumps.
However, jumps do not have a noticeable effect
upon $P(T)$ in the superdroplet simulations we conducted;
see \Fig{fig:pT50mum_SP_theory_1D_Nps}.
By comparing $P(T)$ for $\nps_{\rm back}=40$ (blue crosses in
\Fig{fig:pT50mum_SP_theory_1D_Nps}) with that for $\nps_{\rm back}=2$
(black circles),
while keeping $\nps_{\rm luck}=2$ in both cases,
hardly any jumps occur and the lucky droplet result remains equally accurate.

\begin{figure*}[t]
\begin{center}
\includegraphics[width=\textwidth]{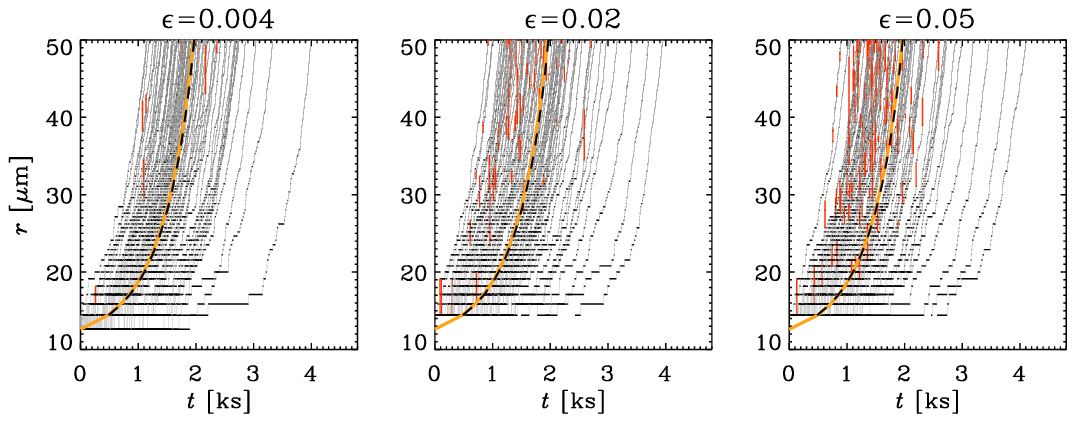}
\end{center}
\caption{\label{fig:ppcollisions}
Growth histories from approach~III for $\epsilon=0.004$ (very few jumps,
relevant to the simulations of \Fig{pT50mum_SP_theory_1D}), as well as
$\epsilon=0.02$, and 0.05, where jumps are more frequent.
The orange thick solid line gives the average collision time and
agrees with that of MFT (thick black \blue{dashed} line) within about a percent.
}
\end{figure*}

For larger values of $\epsilon$, jumps occur much earlier, as can
be seen from \Fig{fig:ppcollisions},
where we show 30 growth curves for the cases $\epsilon=0.004$, which
is relevant to the simulations of \Fig{pT50mum_SP_theory_1D}, as well
as $\epsilon=0.02$, and 0.05.
We also see that for large values of $\epsilon$, the width in
the distribution of arrival times is broader and that both shorter
and longer times are possible.
This suggests that the reason for the finite residual
error in the values of $t_{10\%}$ found by \citet{Dziekan17} for $\nps_i(t_0)>9$
could indeed be due to jumps.
In our superdroplet simulations, by contrast, jumps cannot occur
when $\nps_i(t_0)=1$ or $\nps_{\rm back}\gg \nps_{\rm luck}$.

\subsection{The two aspects of randomness}

Let us now quantify the departure
that is caused by the use of the Monte Carlo collision scheme.
To do this, we need to assess the effects of randomness introduced through
\Eqs{eq:pij}{eq:collision_rate} on the one hand and the random
distribution of the $10\um$ background droplets on the other.
Both aspects enter in the superdroplet algorithm.

We recall that in approach~II, fluctuations originate solely from the random
distribution of the $10\um$ background droplets.
In approach~III, on the other hand, fluctuations
originate solely from the Monte Carlo collision scheme.
By contrast, approach~I is different from either of the two,
because it just uses the exponential distribution of the collision
time intervals, which is indirectly reproduced by the random initial
droplet distribution in approach~II and by the Monte Carlo scheme in
approach~III.

\begin{figure}[t]
\begin{center}
\includegraphics[width=.5\textwidth]{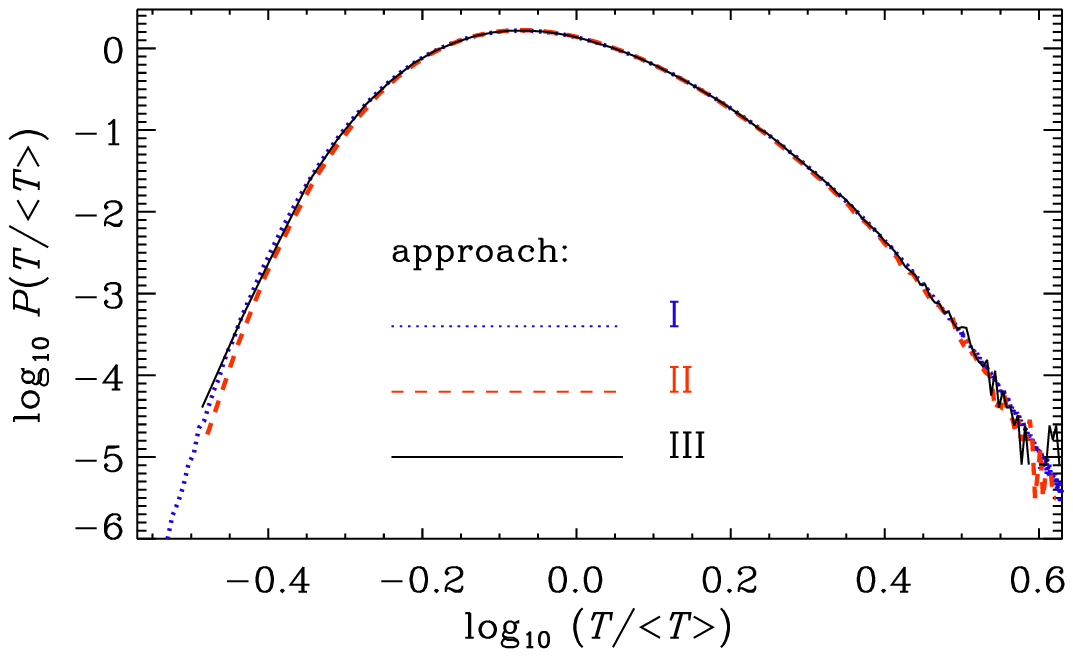}
\end{center}
\caption{\label{fig:pcomp_dens_all_mult_a}
Comparison of $P(T)$ for approaches~I, II, and III.
}
\end{figure}

\begin{table}[t!]
\caption{
Comparison of the moments of $X=\ln(T/\bra{T})$ for approaches I--III.}
\centering
\begin{tabular}{ccccccr}
\hline\hline
Approach & $\bra{X}$ & $\sigma(X)$ & $\SKEW X$ & $\KURT X$ \\
\hline
  I  & $-0.040$ & 0.279 & 0.34 & $ 0.10$ \\
 II  & $-0.039$ & 0.275 & 0.35 & $ 0.11$ \\
III  & $-0.040$ & 0.279 & 0.34 & $ 0.11$ \\
\hline\hline
\end{tabular}
\label{tab:MomentsApproach}
\end{table}

In \Fig{fig:pcomp_dens_all_mult_a}, we compare approaches~I, II, and III.
For our solution using approach~II, we use a nonperiodic domain of
size $10^{-4}\times10^{-4}\times700\m^3$, thus containing on average
2100 droplets.
This was tall enough for the lucky droplet to reach $50\um$ for
all the $10^7$ realizations in this experiment.
The differences between them are very minor, and also the first few
moments are essentially the same; see Table~\ref{tab:MomentsApproach}.
We thus see good agreement between the different approaches.
This suggests that the fluctuations introduced through random
droplet positions is not crucial and that it can be substituted
by the fluctuations of the Monte Carlo scheme alone.

It is worth noting that we were able to
perform $10^7$ and $10^6$ realizations with approaches~II
and III, respectively, and $10^{10}$
realizations with approach~I, while in the superdroplet algorithm
(approach~IV), we could only run $10^3$--$10^4$ realizations due to the
limitation of the computational power.
This may be the reason why fluctuations appear to be slightly
underrepresented in the superdroplet algorithm; see
\Fig{pT50mum_SP_theory_1D} and the discussion in
section~\ref{sec:gh}.\ref{sec:Accuracy}.
Nevertheless, the agreement between the LDM and the superdroplet
simulations demonstrates that the superdroplet algorithm
is able to represent fluctuations during collisions and
does not contain mean-field elements.
This can be further evidenced by the fact that the results of approaches~II and
III agree perfectly with those of approach~I, and the superdroplet algorithm
is just the combination of approaches~II and III.

\section{Discussion}

Fluctuations play a central role in the LDM.
We have therefore used it as a benchmark for our simulation.
It turns out that the superdroplet algorithm is able to reproduce
the growth histories qualitatively and the
distribution of cumulative collision times quantitatively.
The role of fluctuations was also investigated by \citet{Dziekan17},
whose approach to assessing the fluctuations is different from ours.
Instead of analyzing the distribution of cumulative collision times,
as we do here, their primary diagnostics is the time $t_{10\%}$, after
which 10\% of the mass of cloud droplets has reached a radius of $40\um$.
In the LDM, such a time would be infinite, because
there is only one droplet that is allowed to grow.
They then determined the accuracy with which the value of $t_{10\%}$
is determined.
The accuracy increases with the square root of the number of physical
droplets, provided that the ratio $\nps_i(t_0)$ is kept below a limiting value of
about 9.
For $\nps_i(t_0)>9$, they found that there is always a residual error in the
value of $t_{10\%}$ that no longer diminishes as they increase the
number of physical droplets.
We have demonstrated that, when $\nps_i(t_0)>1$, jumps in the growth history
tend to occur.
Those jumps can lead to shorter cumulative collision times, which could
be the source of the residual error they find.

For a given fraction of droplets that first reach a size of $40\um$,
they also determined their average cumulative collision time.
They found a significant dependence on the number of physical droplets.
This is very different in our case where we just have to make sure that
the number of superdroplets is large enough to keep finding collision
partners in the simulations.
However, as the authors point out, this is a consequence of
choosing an initial distribution of droplet sizes that has a finite width.
This implies that for a larger number of droplets, there is a larger
chance that there could be a droplet that is more lucky than for a model
with a smaller number of droplets.
In our case, by contrast, we always have a well-known number of
superdroplets of exactly $12.6\um$, which avoids the sensitivity on the
number of droplets.

The $\nps_i(t_0)=9$ limit of \citet{Dziekan17} does not hold in this
investigation.
In this context we need to recall that their criterion for acceptable
quality concerned the relative error of the time in which 10\%
of the total water has been converted to $40\um$ droplets.
In our case, we have focussed on the shape of the $P(T)$ curve, especially
for small $T$.

\section{Conclusions}
\label{sec:conclusion}

We investigated the growth histories of droplets settling in quiescent air
using superdroplet simulations.
The goal was to determine how accurately these simulations represent
the fluctuations of the growth histories.
This is important because the observed formation time of drizzle-sized
droplets is much
shorter than the one predicted based on the mean collisional cross section. 
The works of \cite{Telford1955}, \cite{Kos05}, and \cite{Wilkinson16}
have shown that this discrepancy can be explained by the presence of stochastic
fluctuations in the time intervals between droplet collisions.
By comparing with the lucky droplet model (LDM) quantitatively, we
have shown that the superdroplet simulations capture the effect of
fluctuations.

A tool to quantify the significance of fluctuations on the
growth history of droplets is the
distribution of cumulative collision times.
Our results show that the superdroplet algorithm
reproduces the distribution of cumulative collision times
that is theoretically expected based on the LDM.
However, the approximation
of representing the dependence of the mean collision
rate on the droplet radius by a power law is not accurate
and must be relaxed for a useful benchmark experiment.

In summary, the superdroplet algorithm appears to take fluctuations fully
into account, at least for the problem of coagulation due to
gravitational settling in quiescent air.
Computing the distribution of cumulative collision times
in the context of turbulent coagulation would be rather
expensive, because one would need to perform many hundreds of fully
resolved 3-D simulations.
Our study suggests that fluctuations are correctly described for
collisions between droplets settling in quiescent fluid, but we do not
know whether this conclusion carries over to the turbulent case.

\acknowledgments
This work was supported through the FRINATEK grant 231444 under the
Research Council of Norway, SeRC, the Swedish Research Council grants 2012-5797, 2013-03992,
and 2017-03865, Formas grant 2014-585, by
the University of Colorado through its support of the
George Ellery Hale visiting faculty appointment,
and by the grant ``Bottlenecks for particle growth in turbulent aerosols''
from the Knut and Alice Wallenberg Foundation, Dnr.\ KAW 2014.0048.
The simulations were performed using resources provided by
the Swedish National Infrastructure for Computing (SNIC)
at the Royal Institute of Technology in Stockholm and
Chalmers Centre for Computational Science and Engineering (C3SE).
This work also benefited from computer resources made available through the
Norwegian NOTUR program, under award NN9405K.

\datastatement
The source code used for the simulations of this study, the {\sc Pencil Code},
is freely available on \url{https://github.com/pencil-code/}.
Datasets for ``Collision fluctuations of lucky droplets with superdroplets''
(v2021.05.07) are available under \blue{\url{https://doi.org/10.5281/zenodo.4742786}}; see also
\url{http://www.nordita.org/~brandenb/projects/lucky/} for easier access.
The plotting and analysis scripts are also included.
Some of the data is stored in the proprietary {\sc idl} save file format.

\appendix

\section{Numerical treatment of approach~I}
\label{AppPencilApproachI}

In section~\ref{sec:CorrectionsKS05}, we noted that solutions to approach~I
have been obtained with the {\sc Pencil Code} \citep{Collaboration2021}.
This might seem somewhat surprising, given that this code is primarily
designed for solving partial differential equations.
It should be realized, however, that this code also provides a
flexible framework for using the message passing interface,
data analysis such as the computation of probability density
distributions, and input/output.

To compute the probability distribution of $T$ with approach~I,
we need to sum up sequences of random numbers for many independent
realizations of $t_k$ drawn from an exponential distribution.
We use the {\tt special/lucky\_droplet} module provided with the code.
Each point in the computational domain corresponds to an
independent realization, so each point is initialized with a
different random seed.
The domain is divided into 1024 smaller domains, allowing
the computational tasks to be performed simultaneously
on 1024 processors, which takes about $4\min$ on a Cray~XC40.

\begin{figure*}[t]\begin{center}
\includegraphics[width=\textwidth]{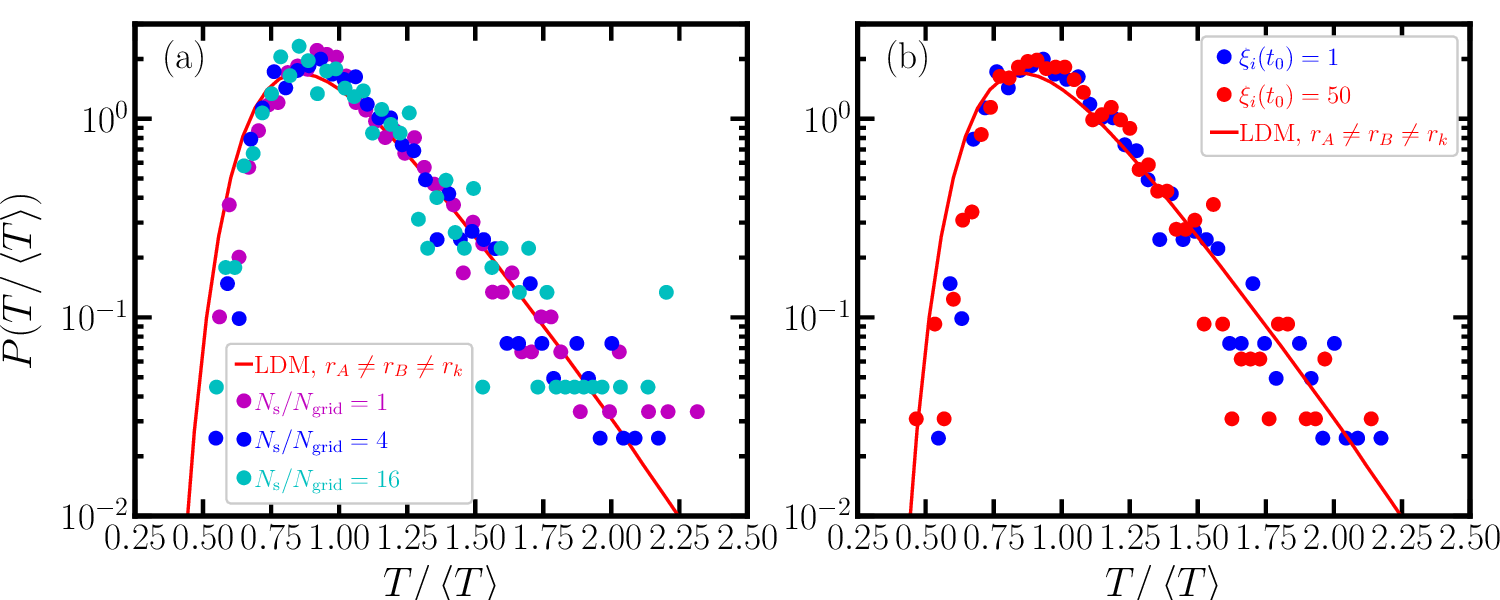}
\end{center}
\caption
{Comparison of $P(T)$ for (a) different $N_{\rm s}/N_{\rm grid}$ with fixed
$\nps_i(t_0)=1$ and (b) for different $\nps_i(t_0)$ with
fixed $N_{\rm s}/N_{\rm grid}=4$.
The blue dots represent $P(T/\langle T\rangle)$ from the
simulation as in \Fig{pT50mum_SP_theory_1D}.
The red curve shows the result for the LDM (approach I) with
$r_{\rm A}\neq r_{\rm B}\neq r_k$,
which is the same simulation as the one in \Fig{pT50mum_SP_theory_1D}.
}
\label{pT50mum_SP_theory_1D_NpNs}
\end{figure*}

\section{Dependence on initial $N_{\rm s}/N_{\rm grid}$ and $N_{\rm d}/N_{\rm s}$}
\label{app:nds}

In this appendix,
we first test the statistical convergence of $P(T)$ for the initial number of
superdroplets per grid cell, $N_{\rm s}(t_0)/N_{\rm grid}$. As discussed
in section~\ref{sec:method}.\ref{subsec:ns}, we set $N_{\rm s}(t_0)/N_{\rm grid}=4$ for 1-D
simulations. Using the same numerical setup,
we examine the statistical convergence of $P(T)$ for different values of
$N_{\rm s}(t_0)/N_{\rm grid}$.
As shown in \Fig{pT50mum_SP_theory_1D_NpNs}(a),
$P(T)$ converges even at $N_{\rm s}(t_0)/N_{\rm grid}=1$.
This is important because one can use as few superdroplets
as possible once $N_{\rm grid}$ is fixed, without suffering from the
statistical fluctuations.

The most practical application of the superdroplet algorithm
is the case when $\xi^i\ge 1$. Thus, we investigate
how $\xi$ affects fluctuations by performing the same 1-D simulation as
described in section~\ref{sec:method}.\ref{subsec:ns} with different values of $\xi^i(t_0)$.
\Fig{pT50mum_SP_theory_1D_NpNs}(b) shows that $P(T)$
is insensitive to $\xi^i(t_0)$, which suggests that the
superdroplet algorithm can capture the effects of fluctuations regardless
of the value of $\xi^i(t_0)$.
This is different from \citet{Dziekan17}, who found that the approach
can represent fluctuations only if $N_{\rm d}(t_0)/N_{\rm s}(t_0)\le9$. 

\section{Horizontal variations of droplet densities}
\label{app:3dLDM}

In this appendix, we analyze in more detail the effect
of horizontal variations of droplet densities discussed
section~\ref{sec:CorrectionsKS05}.
This is relevant for computing the 3-D distribution function
from a 1-D distribution function.
The LDM applies to a given value of the number density.
Other columns have somewhat different number densities and therefore
also different mean cumulative collision times.
The LDM with approaches I--III can be extended to include this
effect by computing cases with different number densities and then
combining $P(T)$ and normalizing by the $\bra{T}$ for the combined $P(T)$.
This can be formulated by introducing the column density as
\begin{equation}
\Sigma(x,y)=\int_{z_1}^{z_2} n(x,y,z)\,{\rm d} z,
\end{equation}
where $z_1$ and $z_2$ denote the vertical slab in which the first
collision occurs, and using this $\Sigma(x,y)$ as a weighting factor
for the 1-D distribution functions $P^{\rm 1D}(T)$ to compute the
3-D distribution functions as
\begin{equation}
P^{\rm 3D}(T)=\left.\int \Sigma(x,y) P^{\rm 1D}(T)\,{\rm d} x\,{\rm d} y
\;\right/\!\!\int \Sigma(x,y)\,{\rm d} x\,{\rm d} y.
\end{equation}
Since the first collision matters the most, we choose
$z_2=z_{\max}$ (where the lucky droplet is released)
and $z_1=z_{\max}-v_2/\lambda_2$ (where it has its first collision).

\begin{table*}[t!]
  \caption{
    Results for approach~II using 30,000 realization
    showing the effects of horizontal density fluctuations in 3-D,
    and comparison with MFT.}
\centering
\begin{tabular}{cc  c  c cc ccc}
\hline\hline
Composition & $\delta n_{\rm rms}/n_0$ & $\delta n_{\max}/n_0$ &  $T_{\min}$ [s] &
$T_{\rm MFT}$ [s] & $\bra{T(n_{\max})}$ [s] & $\bra{T}$ [s] & 
$T_{\min}/\bra{T}$ & $T_{P=0.01}/\bra{T}$ \\ \hline
  (0) &  0   &  0   &  782 & 1969 & 2117 & 2117 & 0.37 & 0.44 \\
  (i) & 0.08 & 0.10 &  795 & 1790 & 1923 & 2126 & 0.37 & 0.42 \\
 (ii) & 0.14 & 0.20 &  767 & 1641 & 1758 & 2155 & 0.36 & 0.40 \\
(iii) & 0.20 & 0.30 &  631 & 1515 & 1628 & 2203 & 0.29 & 0.36 \\
\hline\hline
\end{tabular}
\label{tab:ILPextended}
\end{table*}

Our reference model had a number density of $n_0=10^8\m^{-3}$.
We now consider compositions of models with different values,
where we include the densities
(i) $0.9\times10^8\m^{-3}$ and $1.1\times10^8\m^{-3}$,
as well as (ii) $0.8\times10^8\m^{-3}$ and $1.2\times10^8\m^{-3}$,
and finally also (iii) $0.7\times10^8\m^{-3}$ and $1.3\times10^8\m^{-3}$.
All these compositions have the same mean droplet number
density but different distributions around the mean.
We first average the distribution function and then normalize with
respect to the mean collision time for the ensemble over all columns.
The parameters of the resulting distributions are listed in
Table~\ref{tab:ILPextended} for three compositions with different density
dispersions.
We see that, as we move from composition (i) to compositions (ii) and (iii),
the dispersion ($\delta n_{\rm rms}/n_0$) increases
from 0.08 to 0.14 and 0.20, the distribution
$P(T)$ extends further to both the left and right.
The reference model is listed as (o).
Here we give the rms value of the column-averaged densities, $\bra{n}_i$, as
\begin{equation}
\delta n_{\rm rms}=\left[\sum_{i=0}^{N_i}
\left(\bra{n}_i^2-n_0^2\right)\right]^{1/2},
\end{equation}
where $i$ denotes the column and $N_i$ is the number of columns.
We also give the maximum difference from the average density,
\begin{equation}
\delta n_{\max}=\max_i\left(\bra{n}_i-n_0\right),
\end{equation}
for families (i) with $N_i=2$, (ii) with $N_i=4$, and (iii) with $N_i=6$.
We also list in Table~\ref{tab:ILPextended} several characteristic times
in seconds.
The quantity $T_{\min}$ is the shortest time
in which the lucky droplet reaches $50\um$,
$T_{\rm MFT}$ denotes the value based on MFT,
$\bra{T(n_{\max})}$ is the mean value based on the
column with maximum droplet density and
$\bra{T}$ is the mean based on all columns.
It turns out that for the models of all three families, the value of
$T_{\min}$ agrees with that obtained solely from the model with the highest
density, which is $1.3\times10^8\m^{-3}$ for composition (ii), for example.

The quantity $\bra{T(n_{\max})}$, i.e., the average time for all of
the columns with the largest density, is shorter than the $\bra{T}$
for all the columns, especially for composition (iii) where the largest
densities occur.
For the model (o),
there is only one column, so $\bra{T(n_{\max})}$ is the same as $\bra{T}$.
The value $T_{\rm MFT}$ based on MFT is always somewhat shorter than
$\bra{T(n_{\max})}$.
Finally, we give in Table~\ref{tab:ILPextended} the ratios
$T_{\min}/\bra{T}$ and $T_{P=0.01}/\bra{T}$, where
the subscript $P=0.01$ indicates the argument of $P(T)$ where
the function value is 0.01.


\begin{thebibliography}{53}
\providecommand{\natexlab}[1]{#1}
\providecommand{\url}[1]{\texttt{#1}}
\renewcommand{\UrlFont}{\rmfamily}
\providecommand{\urlprefix}{URL }
\expandafter\ifx\csname urlstyle\endcsname\relax
  \providecommand{\doi}[1]{https://doi.org/\discretionary{}{}{}#1}\else
  \providecommand{\doi}{https://doi.org/\discretionary{}{}{}\begingroup
  \urlstyle{rm}\Url}\fi
\providecommand{\eprint}[2][]{\url{#2}}

\bibitem[{{Arabas} and {Shima}(2013){Arabas}, and {Shima}}]{Arabas13}
{Arabas}, A., and S.~{Shima}, 2013: Large-eddy simulations of trade wind cumuli
  using particle-based microphysics with monte carlo coalescence.
  \textit{Journal of the Atmospheric Sciences}, \textbf{70~(9)}, 2768--2777,
  \doi{10.1175/JAS-D-12-0295.1}.

\bibitem[{Baehr and Klahr(2019)Baehr, and Klahr}]{baehr2019concentration}
Baehr, H., and H.~Klahr, 2019: The concentration and growth of solids in
  fragmenting circumstellar disks. \textit{The Astrophysical Journal},
  \textbf{881~(2)}, 162, \doi{10.3847/1538-4357/ab2f85}.

\bibitem[{Brdar and Seifert(2018)Brdar, and Seifert}]{brdar2018mcsnow}
Brdar, S., and A.~Seifert, 2018: Mcsnow: A monte-carlo particle model for
  riming and aggregation of ice particles in a multidimensional microphysical
  phase space. \textit{Journal of Advances in Modeling Earth Systems},
  \textbf{10~(1)}, 187--206, \doi{10.1002/2017MS001167}.

\bibitem[{{Drakowska} et~al.(2014){Drakowska}, {Windmark},, and
  {Dullemond}}]{Dullemond_2014}
{Drakowska}, J., F.~{Windmark}, and C.~P. {Dullemond}, 2014: {Modeling dust
  growth in protoplanetary disks: The breakthrough case}. \textit{Astronomy \&
  Astrophysics}, \textbf{567}, A38, \doi{10.1051/0004-6361/201423708},
  \eprint{1406.0870}.

\bibitem[{Dziekan and Pawlowska(2017)Dziekan, and Pawlowska}]{Dziekan17}
Dziekan, P., and H.~Pawlowska, 2017: Stochastic coalescence in lagrangian cloud
  microphysics. \textit{Atmospheric Chemistry and Physics}, \textbf{17~(22)},
  13\,509--13\,520, \doi{10.5194/acp-17-13509-2017}.

\bibitem[{Dziekan et~al.(2019)Dziekan, Waruszewski,, and
  Pawlowska}]{dziekan2019university}
Dziekan, P., M.~Waruszewski, and H.~Pawlowska, 2019: University of {Warsaw
  Lagrangian Cloud Model (UWLCM)} 1.0: a modern large-eddy simulation tool for
  warm cloud modeling with {L}agrangian microphysics. \textit{Geoscientific
  Model Development}, \textbf{12~(6)}, 2587--2606,
  \doi{10.5194/gmd-12-2587-2019}.

\bibitem[{Gillespie(1972)}]{gillespie1972stochastic}
Gillespie, D.~T., 1972: The stochastic coalescence model for cloud droplet
  growth. \textit{Journal of the Atmospheric Sciences}, \textbf{29~(8)},
  1496--1510, \doi{10.1175/1520-0469(1972)029<1496:tscmfc>2.0.co;2}.

\bibitem[{{Grabowski}(2020)}]{grabowski2020comparison}
{Grabowski}, W.~W., 2020: {Comparison of Eulerian Bin and Lagrangian
  Particle-Based Microphysics in Simulations of Nonprecipitating Cumulus}.
  \textit{Journal of Atmospheric Sciences}, \textbf{77~(11)}, 3951--3970,
  \doi{10.1175/JAS-D-20-0100.1}.

\bibitem[{Grabowski et~al.(2019)Grabowski, Morrison, Shima, Abade, Dziekan,,
  and Pawlowska}]{grabowski2019modeling}
Grabowski, W.~W., H.~Morrison, S.-I. Shima, G.~C. Abade, P.~Dziekan, and
  H.~Pawlowska, 2019: Modeling of cloud microphysics: Can we do better?
  \textit{Bulletin of the American Meteorological Society}, \textbf{100~(4)},
  655--672, \doi{10.1175/BAMS-D-18-0005.1}.

\bibitem[{Hoffmann et~al.(2019)Hoffmann, Yamaguchi,, and
  Feingold}]{hoffmann2019inhomogeneous}
Hoffmann, F., T.~Yamaguchi, and G.~Feingold, 2019: {Inhomogeneous mixing in
  Lagrangian cloud models: Effects on the production of precipitation embryos}.
  \textit{Journal of the Atmospheric Sciences}, \textbf{76~(1)}, 113--133,
  \doi{10.1175/JAS-D-18-0087.1}.

\bibitem[{Jaruga and Pawlowska(2018)Jaruga, and
  Pawlowska}]{jaruga2018libcloudph}
Jaruga, A., and H.~Pawlowska, 2018: libcloudph++ 2.0: aqueous-phase chemistry
  extension of the particle-based cloud microphysics scheme.
  \textit{Geoscientific Model Development}, \textbf{11~(9)}, 3623--3645,
  \doi{10.5194/gmd-11-3623-2018}.

\bibitem[{Johansen et~al.(2015)Johansen, Mac~Low, Lacerda,, and
  Bizzarro}]{johansen2015growth}
Johansen, A., M.-M. Mac~Low, P.~Lacerda, and M.~Bizzarro, 2015: Growth of
  asteroids, planetary embryos, and kuiper belt objects by chondrule accretion.
  \textit{Science Advances}, \textbf{1~(3)}, e1500\,109,
  \doi{10.1126/sciadv.1500109}.

\bibitem[{{Johansen} et~al.(2012){Johansen}, {Youdin},, and
  {Lithwick}}]{Johansen_2012}
{Johansen}, A., A.~N. {Youdin}, and Y.~{Lithwick}, 2012: {Adding particle
  collisions to the formation of asteroids and Kuiper belt objects via
  streaming instabilities}. \textit{Astron, Astroph.}, \textbf{537}, A125,
  \doi{10.1051/0004-6361/201117701}.

\bibitem[{Kobayashi et~al.(2019)Kobayashi, Isoya,, and
  Sato}]{kobayashi2019importance}
Kobayashi, H., K.~Isoya, and Y.~Sato, 2019: Importance of giant impact ejecta
  for orbits of planets formed during the giant impact era. \textit{The
  Astrophysical Journal}, \textbf{887~(2)}, 226,
  \doi{10.3847/1538-4357/ab5307}.

\bibitem[{Kostinski and Shaw(2005)Kostinski, and Shaw}]{Kos05}
Kostinski, A.~B., and R.~A. Shaw, 2005: Fluctuations and luck in droplet growth
  by coalescence. \textit{Bull. Am. Met. Soc.}, \textbf{86}, 235--244,
  \doi{10.1175/BAMS-86-2-235}.

\bibitem[{Lamb and Verlinde(2011)Lamb, and Verlinde}]{lamb_verlinde_2011}
Lamb, D., and J.~Verlinde, 2011: \textit{Growth by collection}, 380--414.
  Cambridge University Press, \doi{10.1017/CBO9780511976377}.

\bibitem[{Li et~al.(2017)Li, Brandenburg, Haugen,, and Svensson}]{li17}
Li, X.-Y., A.~Brandenburg, N.~E.~L. Haugen, and G.~Svensson, 2017: {Eulerian
  and Lagrangian approaches to multidimensional condensation and collection}.
  \textit{J. Adv. Modeling Earth Systems}, \textbf{9}, 1116--1137,
  \doi{10.1002/2017MS000930}.

\bibitem[{Li et~al.(2018)Li, Brandenburg, Svensson, Haugen, Mehlig,, and
  Rogachevskii}]{li2017effect}
Li, X.-Y., A.~Brandenburg, G.~Svensson, N.~E. Haugen, B.~Mehlig, and
  I.~Rogachevskii, 2018: Effect of turbulence on collisional growth of cloud
  droplets. \textit{Journal of the Atmospheric Sciences}, \textbf{75~(10)},
  3469--3487, \doi{10.1175/JAS-D-18-0081.1}.

\bibitem[{Li et~al.(2020)Li, Brandenburg, Svensson, Haugen, Mehlig,, and
  Rogachevskii}]{li2018condensational}
Li, X.-Y., A.~Brandenburg, G.~Svensson, N.~E. Haugen, B.~Mehlig, and
  I.~Rogachevskii, 2020: Condensational and collisional growth of cloud
  droplets in a turbulent environment. \textit{Journal of the Atmospheric
  Sciences}, \textbf{77~(1)}, 337--353, \doi{10.1175/JAS-D-19-0107.1}.

\bibitem[{Li and Mattsson(2020)Li, and Mattsson}]{li2020dust}
Li, X.-Y., and L.~Mattsson, 2020: Dust growth by accretion of molecules in
  supersonic interstellar turbulence. \textit{The Astrophysical Journal},
  \textbf{903~(2)}, 148, \doi{10.3847/1538-4357/abb9ad}.

\bibitem[{Li and Mattsson(2021)Li, and Mattsson}]{li2020coagulation}
Li, X.-Y., and L.~Mattsson, 2021: Coagulation of inertial particles in
  supersonic turbulence. \textit{Astronomy \& Astrophysics}, \textbf{648}, A52,
  \doi{10.1051/0004-6361/202040068}.

\bibitem[{Li et~al.(2019)Li, Svensson, Brandenburg,, and Haugen}]{li2018cloud}
Li, X.-Y., G.~Svensson, A.~Brandenburg, and N.~E.~L. Haugen, 2019:
  Cloud-droplet growth due to supersaturation fluctuations in stratiform
  clouds. \textit{Atmospheric Chemistry and Physics}, \textbf{19~(1)},
  639--648, \doi{10.5194/acp-19-639-2019}.

\bibitem[{Madival(2018)}]{madival2018stochastic}
Madival, D.~G., 2018: Stochastic growth of cloud droplets by collisions during
  settling. \textit{Theoretical and Computational Fluid Dynamics},
  \textbf{32~(2)}, 235--244, \doi{10.1007/s00162-017-0451-z}.

\bibitem[{Naumann and Seifert(2015)Naumann, and Seifert}]{Naumann15}
Naumann, A.~K., and A.~Seifert, 2015: {A Lagrangian drop model to study warm
  rain microphysical processes in shallow cumulus}. \textit{Journal of Advances
  in Modeling Earth Systems}, \textbf{7~(3)}, 1136--1154,
  \doi{10.1002/2015MS000456}.

\bibitem[{Naumann and Seifert(2016)Naumann, and Seifert}]{Naumann16}
Naumann, A.~K., and A.~Seifert, 2016: Recirculation and growth of raindrops in
  simulated shallow cumulus. \textit{Journal of Advances in Modeling Earth
  Systems}, \textbf{8~(2)}, 520--537, \doi{10.1002/2016MS000631}.

\bibitem[{Nesvorn{\`y} et~al.(2019)Nesvorn{\`y}, Li, Youdin, Simon,, and
  Grundy}]{nesvorny2019trans}
Nesvorn{\`y}, D., R.~Li, A.~N. Youdin, J.~B. Simon, and W.~M. Grundy, 2019:
  {Trans-Neptunian binaries as evidence for planetesimal formation by the
  streaming instability}. \textit{Nature Astronomy}, \textbf{3~(9)}, 808--812,
  \doi{10.1038/s41550-019-0806-z}.

\bibitem[{Onishi et~al.(2015)Onishi, Matsuda,, and Takahashi}]{onishi2015}
Onishi, R., K.~Matsuda, and K.~Takahashi, 2015: Lagrangian tracking simulation
  of droplet growth in turbulence--turbulence enhancement of autoconversion
  rate. \textit{Journal of the Atmospheric Sciences}, \textbf{72~(7)},
  2591--2607, \doi{10.1175/JAS-D-14-0292.1}.

\bibitem[{Ormel et~al.(2009)Ormel, Paszun, Dominik,, and
  Tielens}]{ormel2009dust}
Ormel, C., D.~Paszun, C.~Dominik, and A.~Tielens, 2009: {Dust coagulation and
  fragmentation in molecular clouds-I. How collisions between dust aggregates
  alter the dust size distribution}. \textit{Astronomy \& Astrophysics},
  \textbf{502~(3)}, 845--869, \doi{10.1051/0004-6361/200811158}.

\bibitem[{Paoli et~al.(2004)Paoli, Helie,, and Poinsot}]{paoli2004contrail}
Paoli, R., J.~Helie, and T.~Poinsot, 2004: Contrail formation in aircraft
  wakes. \textit{Journal of Fluid Mechanics}, \textbf{502}, 361--373,
  \doi{10.1017/S0022112003007808}.

\bibitem[{{Pencil Code Collaboration} et~al.(2021)}]{Collaboration2021}
{Pencil Code Collaboration}, and Coauthors, 2021: {The Pencil Code, a modular
  MPI code for partial differential equations and particles: multipurpose and
  multiuser-maintained}. \textit{The Journal of Open Source Software},
  \textbf{6~(58)}, 2807, \doi{10.21105/joss.02807}, \eprint{2009.08231}.

\bibitem[{Poon et~al.(2020)Poon, Nelson, Jacobson,, and
  Morbidelli}]{poon2020formation}
Poon, S.~T., R.~P. Nelson, S.~A. Jacobson, and A.~Morbidelli, 2020: {Formation
  of compact systems of super-Earths via dynamical instabilities and giant
  impacts}. \textit{Monthly Notices of the Royal Astronomical Society},
  \textbf{491~(4)}, 5595--5620, \doi{10.1093/mnras/stz3296}.

\bibitem[{Pruppacher and Klett(1997)Pruppacher, and Klett}]{Pru78}
Pruppacher, H.~R., and J.~D. Klett, 1997: \textit{Microphysics of clouds and
  precipitation, 2nd edition}. Kluwer Academic Publishers, Dordrecht, The
  Nederlands, \doi{10.1080/02786829808965531}, 954p.

\bibitem[{Riechelmann et~al.(2012)Riechelmann, Noh,, and
  Raasch}]{Riechelmann12}
Riechelmann, T., Y.~Noh, and S.~Raasch, 2012: {A new method for large-eddy
  simulations of clouds with Lagrangian droplets including the effects of
  turbulent collision}. \textit{New Journal of Physics}, \textbf{14~(6)},
  065\,008, \doi{10.1088/1367-2630/14/6/065008}.

\bibitem[{Ros and Johansen(2013)Ros, and Johansen}]{ros2013ice}
Ros, K., and A.~Johansen, 2013: Ice condensation as a planet formation
  mechanism. \textit{Astronomy \& Astrophysics}, \textbf{552}, A137,
  \doi{10.1051/0004-6361/201220536}.

\bibitem[{Ros et~al.(2019)Ros, Johansen, Riipinen,, and
  Schlesinger}]{ros2019effect}
Ros, K., A.~Johansen, I.~Riipinen, and D.~Schlesinger, 2019: Effect of
  nucleation on icy pebble growth in protoplanetary discs. \textit{Astronomy \&
  Astrophysics}, \textbf{629}, A65, \doi{10.1051/0004-6361/201834331}.

\bibitem[{Saffman and Turner(1956)Saffman, and Turner}]{1955_Saffman}
Saffman, P.~G., and J.~S. Turner, 1956: On the collision of drops in turbulent
  clouds. \textit{Journal of Fluid Mechanics}, \textbf{1}, 16--30,
  \doi{10.1017/S0022112056000020}.

\bibitem[{Saito and Gotoh(2018)Saito, and Gotoh}]{saito2018turbulence}
Saito, I., and T.~Gotoh, 2018: Turbulence and cloud droplets in cumulus clouds.
  \textit{New Journal of Physics}, \textbf{20~(2)}, 023\,001,
  \doi{10.1088/1367-2630/aaa229}.

\bibitem[{Sato et~al.(2017)Sato, Shima,, and Tomita}]{sato2017grid}
Sato, Y., S.-i. Shima, and H.~Tomita, 2017: {A grid refinement study of trade
  wind cumuli simulated by a Lagrangian cloud microphysical model: the
  super-droplet method}. \textit{Atmospheric Science Letters}, \textbf{18~(9)},
  359--365, \doi{10.1002/asl.764}.

\bibitem[{Sato et~al.(2018)Sato, Shima,, and Tomita}]{sato2018numerical}
Sato, Y., S.-i. Shima, and H.~Tomita, 2018: {Numerical convergence of shallow
  convection cloud field simulations: Comparison between double-moment Eulerian
  and particle-based Lagrangian microphysics coupled to the same dynamical
  core}. \textit{Journal of Advances in Modeling Earth Systems},
  \textbf{10~(7)}, 1495--1512, \doi{10.1029/2018MS001285}.

\bibitem[{Seifert et~al.(2019)Seifert, Leinonen, Siewert,, and
  Kneifel}]{seifert2019geometry}
Seifert, A., J.~Leinonen, C.~Siewert, and S.~Kneifel, 2019: The geometry of
  rimed aggregate snowflakes: A modeling study. \textit{Journal of Advances in
  Modeling Earth Systems}, \textbf{11~(3)}, 712--731,
  \doi{10.1029/2018MS001519}.

\bibitem[{{Shima} et~al.(2009){Shima}, {Kusano}, {Kawano}, {Sugiyama},, and
  {Kawahara}}]{Shima09}
{Shima}, S., K.~{Kusano}, A.~{Kawano}, T.~{Sugiyama}, and S.~{Kawahara}, 2009:
  {The super-droplet method for the numerical simulation of clouds and
  precipitation: a particle-based and probabilistic microphysics model coupled
  with a non-hydrostatic model}. \textit{Quart. J. Roy. Met. Soc.},
  \textbf{135}, 1307--1320, \doi{10.1002/qj.441}, \eprint{physics/0701103}.

\bibitem[{Shima et~al.(2020)Shima, Sato, Hashimoto,, and
  Misumi}]{shima2020predicting}
Shima, S.-i., Y.~Sato, A.~Hashimoto, and R.~Misumi, 2020: {Predicting the
  morphology of ice particles in deep convection using the super-droplet
  method: development and evaluation of SCALE-SDM 0.2. 5-2.2. 0,-2.2. 1,
  and-2.2. 2}. \textit{Geoscientific Model Development}, \textbf{13~(9)},
  4107--4157, \doi{10.5194/gmd-13-4107-2020}.

\bibitem[{{Sokal}(1997)}]{Sok97}
{Sokal}, A., 1997: \textit{{Monte Carlo Methods in Statistical Mechanics:
  Foundations and New Algorithms}}. Boston: Springer,
  \doi{10.1007/978-1-4899-0319-8_6}.

\bibitem[{S{\"o}lch and K{\"a}rcher(2010)S{\"o}lch, and
  K{\"a}rcher}]{solch2010large}
S{\"o}lch, I., and B.~K{\"a}rcher, 2010: {A large-eddy model for cirrus clouds
  with explicit aerosol and ice microphysics and Lagrangian ice particle
  tracking}. \textit{Quarterly Journal of the Royal Meteorological Society},
  \textbf{136~(653)}, 2074--2093, \doi{10.1002/qj.689}.

\bibitem[{Telford(1955)}]{Telford1955}
Telford, J.~W., 1955: A new aspect of coalescence theory. \textit{Journal of
  Meteorology}, \textbf{12~(5)}, 436--444,
  \doi{10.1175/1520-0469(1955)012<0436:anaoct>2.0.co;2}.

\bibitem[{Twomey(1964)}]{twomey1964statistical}
Twomey, S., 1964: Statistical effects in the evolution of a distribution of
  cloud droplets by coalescence. \textit{Journal of the Atmospheric Sciences},
  \textbf{21~(5)}, 553--557,
  \doi{10.1175/1520-0469(1964)021<0553:seiteo>2.0.co;2}.

\bibitem[{Unterstrasser et~al.(2017)Unterstrasser, Hoffmann,, and
  Lerch}]{Unterstrasser17}
Unterstrasser, S., F.~Hoffmann, and M.~Lerch, 2017: {Collection/aggregation
  algorithms in Lagrangian cloud microphysical models: rigorous evaluation in
  box model simulations}. \textit{Geoscientific Model Development},
  \textbf{10~(4)}, 1521--1548, \doi{10.5194/gmd-10-1521-2017}.

\bibitem[{Unterstrasser et~al.(2020)Unterstrasser, Hoffmann,, and
  Lerch}]{unterstrasser2020collisional}
Unterstrasser, S., F.~Hoffmann, and M.~Lerch, 2020: {Collisional growth in a
  particle-based cloud microphysical model: insights from column model
  simulations using LCM1D (v1. 0)}. \textit{Geoscientific Model Development},
  \textbf{13~(11)}, 5119--5145, \doi{10.5194/gmd-13-5119-2020}.

\bibitem[{Wilkinson(2016)}]{Wilkinson16}
Wilkinson, M., 2016: Large deviation analysis of rapid onset of rain showers.
  \textit{Physics Review Letters}, \textbf{116}, 018\,501,
  \doi{10.1103/PhysRevLett.116.018501}.

\bibitem[{Yang and Zhu(2020)Yang, and Zhu}]{yang2020morphological}
Yang, C.-C., and Z.~Zhu, 2020: Morphological signatures induced by dust back
  reaction in discs with an embedded planet. \textit{Monthly Notices of the
  Royal Astronomical Society}, \textbf{491~(4)}, 4702--4718,
  \doi{10.1093/mnras/stz3232}.

\bibitem[{Zannetti(1984)}]{zannetti1984new}
Zannetti, P., 1984: New monte carlo scheme for simulating lagranian particle
  diffusion with wind shear effects. \textit{Applied Mathematical Modelling},
  \textbf{8~(3)}, 188--192, \doi{https://doi.org/10.1016/0307-904X(84)90088-X}.

\bibitem[{{Zsom} and {Dullemond}(2008){Zsom}, and {Dullemond}}]{Dullemond_2008}
{Zsom}, A., and C.~P. {Dullemond}, 2008: A representative particle approach to
  coagulation and fragmentation of dust aggregates and fluid droplets.
  \textit{Astronomy \& Astrophysics}, \textbf{489~(2)}, 931--941,
  \doi{10.1051/0004-6361:200809921}.

\bibitem[{Zsom et~al.(2010)Zsom, Ormel, G{\"u}ttler, Blum,, and
  Dullemond}]{zsom2010outcome}
Zsom, A., C.~Ormel, C.~G{\"u}ttler, J.~Blum, and C.~Dullemond, 2010: {The
  outcome of protoplanetary dust growth: pebbles, boulders, or
  planetesimals?-II. Introducing the bouncing barrier}. \textit{Astronomy \&
  Astrophysics}, \textbf{513}, A57, \doi{10.1051/0004-6361/201116515}.

\end{thebibliography}
\end{document}